\documentclass[12pt,a4paper,english]{article}
\usepackage[utf8]{inputenc}
\usepackage{algorithmic}
\usepackage{algorithm}
\usepackage{float}

\usepackage[english]{babel}

\usepackage{graphicx}
\usepackage{epsfig}            
\usepackage{pstricks}          

\usepackage{fancyhdr}
\usepackage[verbose,a4paper,tmargin=30mm,bmargin=37.5mm,lmargin=30mm,rmargin=30mm]{geometry}

\usepackage{helvet}

\usepackage{mathptmx} 
\usepackage[T1]{fontenc}

\usepackage{rotating}

\usepackage{amssymb}       
\usepackage{amsmath}       

\usepackage{natbib} 

\usepackage{multirow}

\usepackage{verbatim}          
\usepackage{subcaption}
\definecolor{METblue}{cmyk}{0.85,0,0.2,0.2}

\usepackage{titlesec}

\usepackage{url}

\titleformat{\section}
  [block]
  { \normalfont\sffamily\Large\bfseries\color{METblue} }
  {\makebox[2em][r]{\thesection}}
  {5mm}
  {\vspace{5mm}}

\titleformat{\subsection}[block]%
  {\normalfont\sffamily\bfseries}%
  {\makebox[2em][r]{\thesubsection}}%
  {5mm}
  {\vspace{3mm}}[]
\titleformat{\subsubsection}[block]%
  {\normalfont\sffamily\bfseries}%
  {\makebox[3em][r]{\thesubsubsection}}%
  {5mm}
  {\vspace{3mm}}[]

\titlespacing*{\section}{-17mm}{5mm}{0mm}
\titlespacing*{\subsection}{-13.5mm}{5mm}{0mm}
\titlespacing*{\subsubsection}{-17.5mm}{5mm}{0mm}

\usepackage{tocloft}

\begin{document}

\bibliographystyle{agufull04}
 
\thispagestyle{empty}  

\noindent
\begin{tabular}{@{} p{63mm} p{50mm} r}
\includegraphics*[]{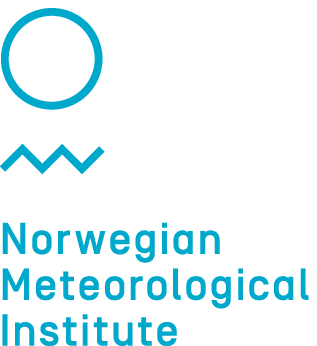} 
&
\fontsize{27.5pt}{33pt} \selectfont \bf \sffamily MET{\color{gray} report}
&
 \begin{minipage}[b]{28mm}
  \begin{flushright}
   \footnotesize \sffamily No. 07/2020 \\ ISSN 2387-4201 \\ Free              
  \end{flushright}
 \end{minipage}
\end{tabular}

\vfill

\begin{flushright}
{\fontsize{20pt}{25pt}\selectfont \bf \sffamily \emph{seNorge} observational gridded datasets}          

\vspace{2mm}
{\fontsize{12.5pt}{15pt}\selectfont \sffamily seNorge\_2018, version 20.05 
\\
}
\end{flushright}
{\fontsize{12.5pt}{15pt}\selectfont \sffamily Cristian Lussana \\  }                                  
{\fontsize{9pt}{11pt}\selectfont \sffamily The Norwegian Meteorological Institute, Oslo, Norway}
\vspace{3mm}

\begin{figure}[!h]
\begin{center}
\includegraphics[width=14cm,height=11.5cm]{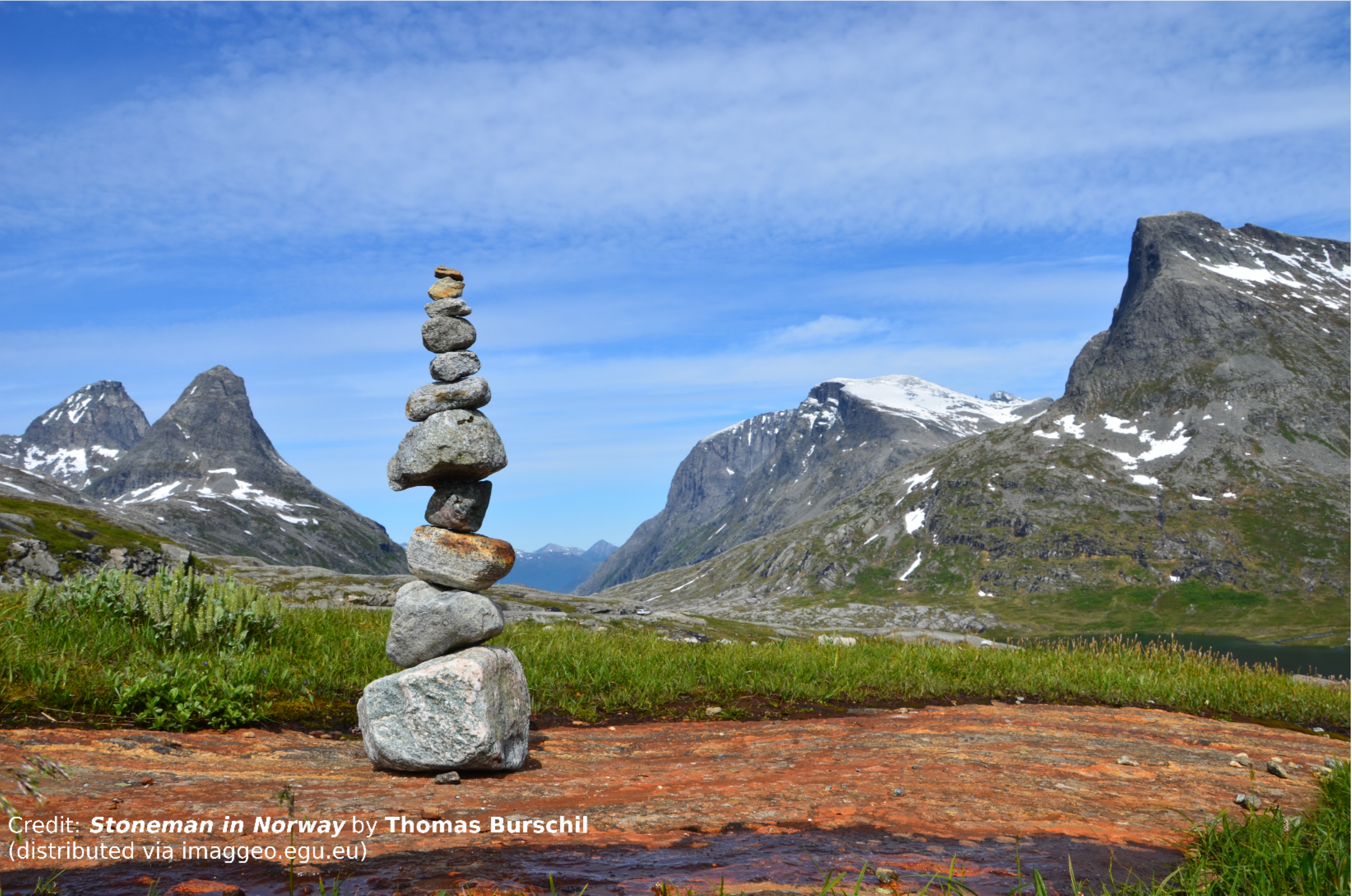}          
\end{center}
\end{figure}



\clearpage

\thispagestyle{fancy} 
\headheight=15pt
\renewcommand{\headrulewidth}{0pt}

\setlength{\unitlength}{1mm}  

\begin{table}[!ht]

\begin{tabular}[c]{lr}
\vspace{5mm}
\includegraphics*{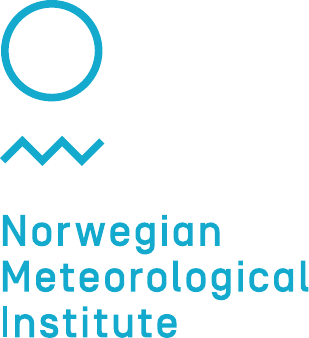} & \hspace{43mm}
{\fontsize{27.5pt}{33pt}\selectfont \bf \sffamily MET{\color{gray} report}}\\
\end{tabular}

\sffamily{
\begin{tabular}[t]{|p{110mm}|p{40mm}|} \hline
{\bf \sffamily Title }                  & {\bf \sffamily Date}               \\ 
seNorge observational gridded dataset. seNorge\_2018, version 20.05.
                             & \today                   \\ \hline
{\bf \sffamily Section}                & {\bf \sffamily Report no.}         \\ 
Division for Climate Services                        &  07/2020                  \\ \hline
{\bf \sffamily Author(s)}                 & {\bf \sffamily Classification}     \\ 
Cristian Lussana              
                             & \begin{picture}(20,4)(-2,-1.0)
                               \put (0,0){\circle*{4}}
                               \put (7,0){\makebox(0,0){Free}}
                               \put (15,0){\circle{4}}
                               \put (27,0){\makebox(0,0){Restricted}}
                               \end{picture}
                               \\ \hline
\end{tabular}

\begin{tabular}[t]{|p{154.3mm}|}
{\bf \sffamily Abstract}                                          \\
seNorge\_2018 is an observational gridded dataset for daily aggregated temperatures and precipitation data over the Norwegian mainland.
The time interval covered spans more than 60 years, from the 1st of January 1957 to the present day.
Version 20.05 of the historical archive (1957-2019) is described in this document, with special emphasis on the variations in the methods with respect to the previous version of the archive, which is version 18.12.
\\
\hline
{\bf \sffamily Keywords}                                          \\ 
  observational gridded datasets, temperature, precipitation, climatology, hydrology, statistical interpolation  \\ 
\hline
\end{tabular}
}

\begin{tabular}[t]{cc}
                             &                            \\
                             &                            \\
                             &                            \\
\line(1,0){70}               & \line(1,0){70}             \\ 
Disciplinary signature       & Responsible signature      \\
Hans Olav Hygen              &  Cecilie Stenersen         \\       
\hspace{75mm}                & \hspace{75mm}              \\

\end{tabular}
\end{table}

\fancyfoot{
\begin{tabular}[b]{p{40mm}p{25mm}p{25mm}p{25mm}p{25mm}}
 \begin{minipage}[l]{40mm} \tiny \color{METblue} {\bf Norwegian Meteorological Institute}\\ Org.no 971274042\\ post@met.no\\ www.met.no / www.yr.no
 \end{minipage} & 
 \begin{minipage}[l]{25mm} \tiny \color{METblue} {\bf Oslo}\\ P.O. Box 43, Blindern\\ 0313 Oslo, Norway\\ T. +47 22 96 30 00
 \end{minipage} &
 \begin{minipage}[l]{25mm} \tiny \color{METblue} {\bf Bergen}\\ All\'egaten 70\\ 5007 Bergen, Norway\\ T. +47 55 23 66 00
 \end{minipage} & 
 \begin{minipage}[l]{25mm} \tiny \color{METblue} {\bf Troms\o}\\ P.O. Box 6314, Langnes\\ 9293 Troms\o, Norway\\ T. +47 77 62 13 00
 \end{minipage} & 
 \begin{minipage}[l]{25mm} \tiny \color{METblue} 
 \end{minipage}
\end{tabular}
}

\clearpage
\tableofcontents
\clearpage

\section{Introduction \label{sec:intro}}
\noindent seNorge indicates a family of observational gridded datasets over Norway. 
The Norwegian Meteorological Institute (MET Norway) started to produce the seNorge datasets versions 1.0 and 1.1 for the Norwegian mainland in the nineties \citep{tveito1999mapping,tveito2000nordic,tveito2002,tveito2005gis,mohr2008new}. 
The version 2.0 has been produced after 2015 \citep{lussana2018qjrms,lussana2018essd}.
The most recent member of this family is seNorge\_2018 and this document describes version 20.05.
In particular, we focus on the modifications of methods and data used with respect to version 18.12, which has been documented by \cite{lussana2019essd}.

The main motivation behind the development of seNorge is to provide gridded datasets of a few near-surface atmospheric key-variables for climatological and hydrological applications.
Observational gridded datasets exploit the information provided by networks of traditional weather stations over a region and they return estimates of the selected variables potentially anywhere within the region.

The tag ''2018'' in seNorge\_2018 indicates that the statistical models used for spatial analysis and the code have been developed in 2018.
An important component of the spatial analysis procedure is Optimal Interpolation \citep[OI][]{gandin1965objective}, and the reader may refer to \cite{ULS2008} for the description of the modified OI scheme used.
The versioning of seNorge\_2018 indicates the production date of the datasets, such that version 20.05 (or ver. 20.05) refers to May 2020.
As stated above, the first version of seNorge\_2018 was ver. 18.12.
A new version may include both updated data sources and further developments of the methods.
Note that the new developments are not considered significant enough to change the name of seNorge\_2018, because the Bayesian inference behind the statistical methods remains unchanged.

seNorge\_2018 includes gridded fields of daily mean, maximum and minimum temperature, and daily total precipitation amounts.
The time period covered by the datasets begins the 1st of January 1957 (1957-01-01) and continues to the present day.
The dataset is updated on a daily basis.
seNorge\_2018 datasets can be divided into two archives: an historical archive and an operational archive. 
The historical archive includes data from 1957-01-01 to a date in the recent past, which is 2019-12-31 for ver. 20.05.
The historical archive is re-built periodically. Then, a new version of the archive is created and new Digital Object Identifiers (DOIs) are assigned to the corresponding datasets.
The operational archive is updated every day based on the data of the last few days.
In addition, each time a new historical archive is built, the operational archive is also modified, such that there is a seamless transition between the most recent version of the historical archive and the operational archive.
In this sense, the operational archive is a provisional archive.
In the case of ver. 20.05, the operational archive starts at 2020-01-01.

The motivations that led us to change parts of the methods used for spatial analysis in ver. 20.05 with respect to ver. 18.12 are different for temperature and precipitation.
In the case of precipitations, the gridded fields of a regional climate model are used as input data to scale precipitation in the spatial analysis. 
Since a climate model covering the same region but spanning a longer time period is available, the gridded fields from this model are used to replace those previously used for ver. 18.12.
In the case of temperature, the OI in ver. 18.12 was configured to achieve the best agreement -on average- between (independent) observations and cross-validation analyses.
However, the optimization procedure favoured some regions and penalized others, depending on the spatial distributions of the observational network.
In particular, since most of the stations used are located at low elevations, the optimization procedure was biased towards ensuring the optimal performances for those regions, while the uncertainties in mountainous regions and in data sparse regions was considered only marginally in the optimization.
As a consequence, the variability of ver. 18.12 analyses fields in data sparse regions is rather large, larger than in seNorge2 (or seNorge version 2.0) for instance.
The ideal situation would be to rely on the OI performances of seNorge\_2018 ver. 18.12 in data dense regions and, at the same time, reduce the uncertainty of the temperature fields in data sparse regions down to the level of seNorge2.
seNorge\_2018 ver. 20.05 has exactly this objective.
The variability of the gridded analysis field in data sparse regions is reduced by blending together the results of several different OIs based on slightly different configurations.

The document is organized as follows.
Sec.~\ref{sec:data} defines seNorge\_2018 variables and presents the data used.
Sec.~\ref{sec:methods} describes the variations in the spatial analysis with respect to the methods implemented for ver. 18.12.
Sec.~\ref{sec:results} compares the gridded fields of ver. 20.05 with those of ver. 18.12 with the aim of studying the effects of variations in the input data and methods over the results.
Sec.~\ref{sec:conclusions} summarizes the main messages of the document.
The data access is then described in the Appendix. 
After the Appendix, a section with supplementary material is included, in order to make the reading of the main text more fluent by avoiding the interruptions caused by pages and pages of figures.

\section{Data and Definitions \label{sec:data}}
\noindent
seNorge covers the Norwegian mainland and it includes all the catchments of rivers flowing in the Norwegian territory, as a consequence seNorge extends also into parts of Sweden and Finland bordering Norway.
The domain of seNorge\_2018 ver. 20.05 is the same of ver. 18.12. 
More details and figures on the domain can be found in previous papers, such as \cite{lussana2018qjrms,lussana2018essd,lussana2019essd}.
The gridded data are presented on a high-resolution terrain-following grid with 1 km spacing.
The coordinate reference system is Universal Transverse Mercator, zone 33.

The data sources used are: MET Norway's climatological archive (frost.met.no), which includes also observations from stations managed by several public Norwegian institutions, such as The Norwegian Water Resources and Energy Directorate (NVE), the Norwegian Public Roads Administration (Statens vegvesen, SVV) and the Norwegian Institute of Bioeconomy Research (NIBIO); the Swedish Meteorological and Hydrological Institute (SMHI) climatological archive; Finnish Meteorological Institute (FMI) climatological archive.
In addition, the daily datasets provided by the European Climate Assessment \& Dataset project \citep[ECA\&D, https://www.ecad.eu,][]{klein2002daily} are also used.

The definitions of the variables available in seNorge\_2018 are:
\begin{itemize}
    \item RR at a given date is the total amount of precipitation accumulated from 06 UTC of the day previous to that date to 06 UTC of the day in the date. RR is also referred to as daily precipitation. Note that RR at a station location does not correspond to the value observed by the corresponding ombrometer, not only because of the adjustment made by the spatial analysis but also because observations are adjusted for the wind-induced under-catch according to the relationships reported by \cite{lussana2019essd} and based on \cite{wolff2015derivation}.
    \item TG at a given date is the mean averaged temperature from 06 UTC of the day previous to that date to 06 UTC of the day in the date. TG is also referred to as daily mean temperature. In detail, it is the arithmetic mean of 24 hourly values or a formula based mean value computed from fewer observations \citep{foerland_tveito1997}.
    \item TX at a given date is the maximum temperature from 18 UTC of the day previous to that date to 18 UTC of the day in the date. TX is also referred to as daily maximum temperature.
    \item TN at a given date is the minimum temperature from 18 UTC of the day previous to that date to 18 UTC of the day in the date. TX is also referred to as daily minimum temperature.
\end{itemize}

\begin{figure}[h!]
    \includegraphics[width=\textwidth]{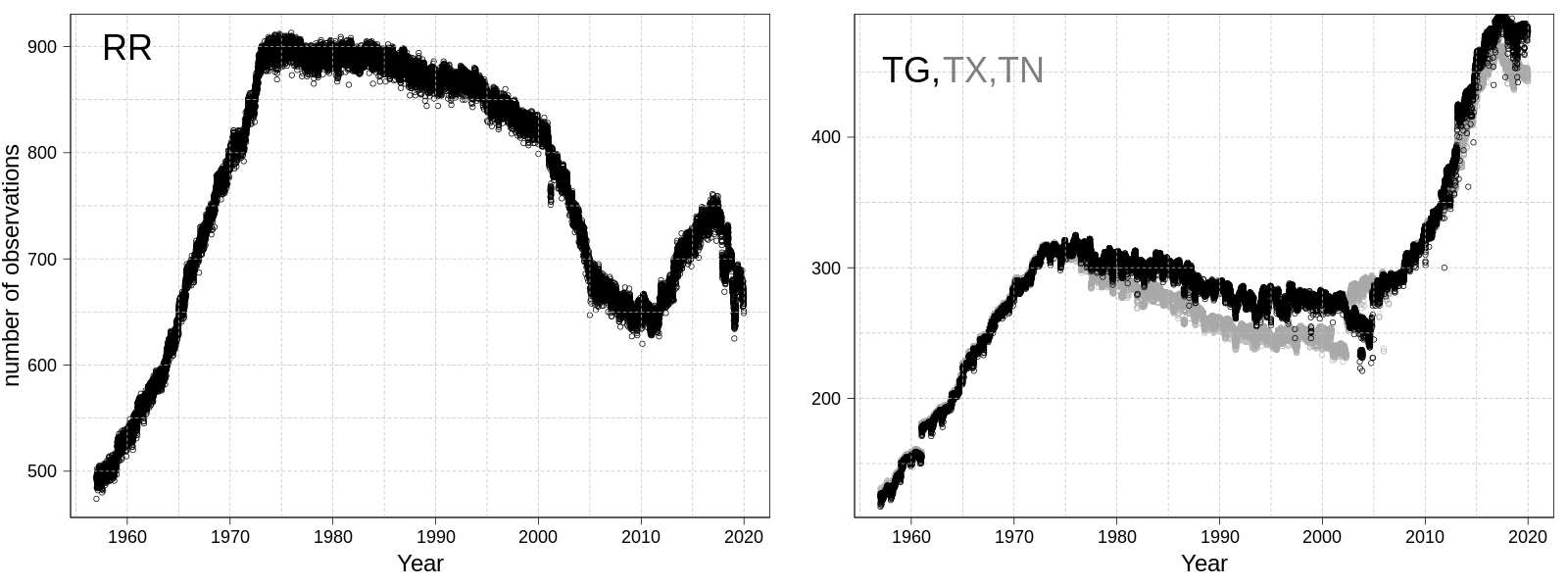}
    \caption{Time series of number of observations used for spatial analysis that are in the seNorge\_2018 domain after the data quality control. The left panel shows RR observations. The right panel shows TG (black dots) and TX, TN (gray dots) observations. Note the different scales on the y-axes between the two panels.}
    \label{fig:data_avail}
\end{figure}

The 63-year time period covered by the historical archive of seNorge\_2018 ver. 20.05 ranges from January 1957 to December 2019.
In such a long period of time the observational network has continuously changed.
The data availability is shown in Fig.~\ref{fig:data_avail}.
There are more observations for precipitation than for temperature.
In the case of temperature, there are no significant differences between TG, TX and TN.
The area of seNorge\_2018 spatial domain measures $606579 \; \mathrm{km}^2$.
In the idealized case of a completely uniform spatial distribution of the observational network, the ratio (area) / (number of observations) corresponds to the area around each observation where that observation is the closest data source.
In this sense, that area can be referred to as a sort of ''area of influence'' of an observation.
The radius of the circle equivalent to the area of influence is a measure of the average distance we need to travel in every direction before meeting another observation.
This radius is a rough indicator of the observation spatial density.
Suppose the number of available observations is 100, then the radius is equal to $44$ km.
With numbers of available observations equal to 200, 300, 600 and 900, then we get circle radii of, respectively: $31$ km, $25$ km, $18$ km and $15$ km.
For all variables, it is possible to identify three different regimes in the time series.
From 1957 to 1975, there was an increase in the number of available observations.
From 1975 to 2000-2005, there was a gradual, slow decrease in the number of observations.
The decrease is rather sharp for RR in the period from 2000 to 2005.
From 2005 to 2020, the situation is different between the two quantities.
In the case of precipitation, there is an oscillatory trend with an average of approximately 700 observations, while for temperature a sharp increase from 300 to 500 observations is shown.

The in-situ data shown in Fig.~\ref{fig:data_avail} has been quality controlled according to MET Norway's quality assurance system specifications.
In addition, for temperature, a spatial consistency test such as those described by \cite{LUS2010} has been applied.
For precipitation, the time series have been visually inspected to identify anomalies with respect to neighbouring stations.

By comparing Fig.~\ref{fig:data_avail} with Fig.~1 of \cite{lussana2019essd}, it can be noted that on average ver. 20.05 is based on more observations than ver. 18.12.
However, similarly to ver. 18.12, the in-situ data used in ver. 20.05 are not evenly distributed across the altitude range.
More stations are located in the lowlands, while in the mountains the observational network is quite sparse.
This should have an influence on the uncertainties of the spatially interpolated fields of ver. 20.05 similar to what reported for ver. 18.12 and documented by \cite{lussana2019essd}.

\section{Methods \label{sec:methods}}
\noindent The steps of the procedure used for the production of daily gridded datasets for an arbitrary day are listed in the Algorithm~\ref{a1}.
The mathematical notation and the symbols used are defined in Tab.~\ref{tab:not}.
The objective of this Section is to describe the modifications in the methods of seNorge\_2018 ver. 20.05 compared to ver. 18.12, while the general description of the methods documented by \cite{lussana2019essd} is still the reference for seNorge\_2018.

In the Algorithm~\ref{a1}, we made it clear that a number of data are required for the production of seNorge\_2018.
The fixed datasets used are not reported in the algorithm and those are the digital elevation model and the land area fraction over the 1 km regular grid.
The gridded fields of wind speed are derived from numerical model output, in this way it is possible to associate a value of wind speed to each observation, independently of the actual availability of an observed value of wind speed.
In fact, observed wind speed values would have been more accurate data but we would have inevitably lost some of the precious observations of temperature and precipitation.
We estimated that data loss would have been a worse solution than the less accuracy for the wind speed.
The numerical models used are described in the paper \cite{lussana2019essd} and they are based on MET Norway's products \citep{reistad2011high, muller2017arome, frogner2019}.

\subsection{Precipitation \label{sec:methods_prec}}
\noindent seNorge\_2018 statistical interpolation of precipitation is based on an iterative OI scheme that uses the relative anomalies between observed and reference precipitation values.
The references are long-term monthly averages of precipitation obtained through numerical models.
In the case of ver. 18.12, the reference datasets used for precipitation are based on hourly precipitation provided by the climate model version of HARMONIE (version cy38h1.2), a seamless NWP model framework developed and used by several national meteorological services.
For ver. 20.05, the reference dataset has changed.
The long-term monthly averages of precipitation have been derived by the "NorCP" simulation, which provides high-resolution (3 km) climate model data covering the 21-year time period from 1998-2018 by an updated climate model version of HARMONIE \citep{Lind2020HCLIM}.

\begin{figure}[h!]
    \includegraphics[width=\textwidth]{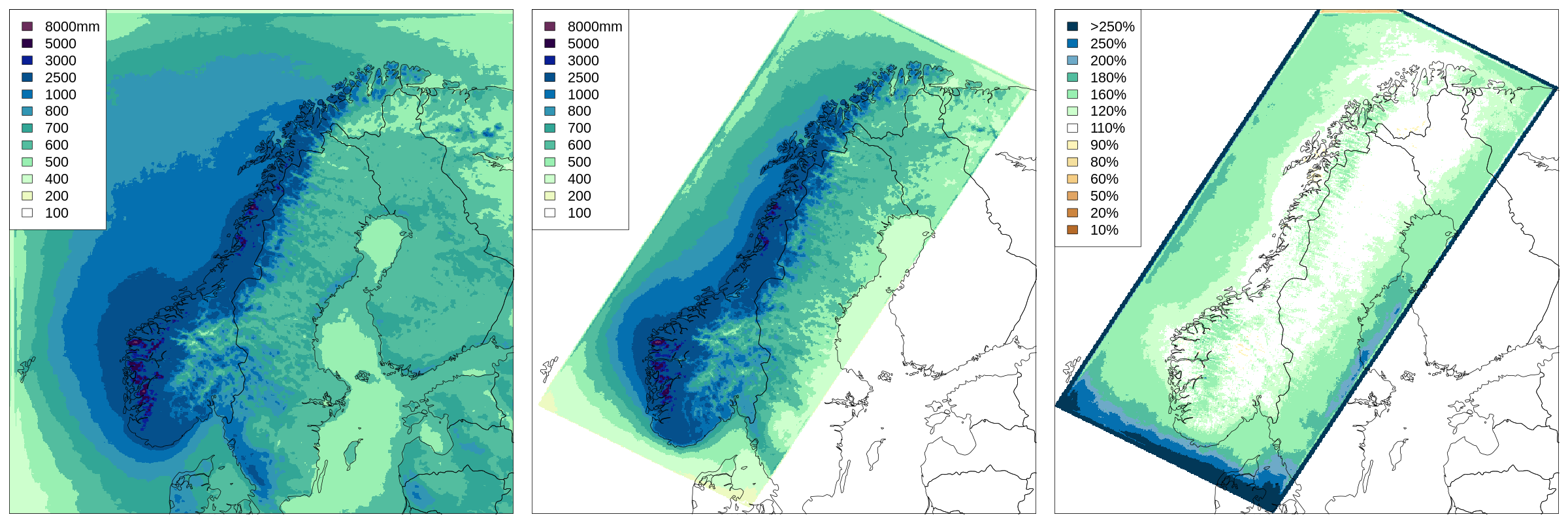}
    \caption{Comparison of typical annual total precipitation fields of the seNorge\_2018 reference fields.
    Left panel: ver. 20.05 reference field.
    Middle panel: ver. 18.12 reference field.
    Right panel: relative deviations as percentages of the ver. 18.12 reference field (i.e., ref(ver 20.05) / ref(ver 18.12)).}
    \label{fig:rr_ref_comp_year}
\end{figure}

In Fig.~\ref{fig:rr_ref_comp_year}, the typical annual total precipitation fields for the references of ver. 20.05 and ver. 18.12 are compared.
First thing to notice is that the domain of the reference for ver. 20.05 is wider than that of ver. 18.12. 
As a consequence, the Norwegian mainland in ver. 20.05 reference is less likely to be affected by those border effects that may affect ver. 18.12, especially in Southern Norway.
Besides, the wider domain of ver. 20.05 should also allow for a better development of the high-resolution dynamics of the dynamical downscaling procedure.
As shown in the right panel, the reference of ver. 20.05 has -on average- larger values of precipitation, though the situation can vary significantly between months and regions, as shown in Figs.~\ref{fig:rr_ref_comp_month01}-~\ref{fig:rr_ref_comp_month12}.

\subsection{Temperatures \label{sec:methods_temp}}
\noindent seNorge\_2018 statistical interpolations of temperature are all based on a two-step spatial analysis scheme.
The first step aims at representing the temperature field at a point as the average over a spatial support larger than the local station density, where this is characterized as the average distance between a station and its closest neighbouring stations in the surroundings of that point.
The second step refines the temperature representation at a point by including those effects that are observed only by a small number of neighbouring stations.
The first step is implemented through the elaboration of sub-regional pseudo-background fields.
This part has remained unchanged from ver. 18.12.
The second step is implemented by means of an Optimal Interpolation (OI) scheme and this part has been significantly changed, as it is described in the following of this Section.

For each of the three variables TG, TX and TN, the OI runs with more than a single configuration, as it is instead for ver. 18.12.
In particular, four different OI are performed.
The parameter that is varying within the different OI runs is $D^z$, which is the vertical decorrelation length used in the definition of the background error covariance matrix.
For ver. 18.12, $D^z$ was set to 210 m.
In the case of ver. 20.05 the four values used are: 210 m; 400 m; 600 m and 800 m.
For each combination of the pair variable/OI configuration, a dataset of gridded fields is produced.

There are two more steps that must be performed in order to obtain the final dataset of TG, TX and TN.
The first step is the merging of the four datasets into a single dataset for each variable.
The second step is the check for consistency between the three variables, specifically to ensure that TX is greater or equal to TG and, at the same time, TN is smaller or equal to TG.
Note that such a check is needed, because the spatial analysis is performed for each variable independently from the others, therefore in regions where the analysis uncertainty at a given time is greater than the difference between variables, the consistency between TG, TX and TN may be violated.
The check for consistency was implemented already in ver. 18.12 and it first replaces TX with TG for those grid points where TG is greater than TX, then it replaces TN with TG for those grid points where TG is smaller than TX.
This simple and practical solution might not always yield satisfactory results, for this reason it is one of the points that should be improved in future versions of seNorge\_2018.

The merging of datasets into a single one is a new feature of ver. 20.05 compared to ver. 18.12. 
Statistical interpolation schemes produce fields where each value is an average over a spatial support and the sizes of those spatial supports vary across the domain.
The OI settings influence the sizes of the three-dimensional spatial supports.
The vector of OI settings $\boldsymbol{\lambda}_1$ includes $D^z=210\,\mathrm{m}$ (same as ver. 18.12), then: $\boldsymbol{\lambda}_2$ includes $D^z=400\,\mathrm{m}$; $\boldsymbol{\lambda}_3$ includes $D^z=600\,\mathrm{m}$ (same as seNorge2); $\boldsymbol{\lambda}_4$ includes $D^z=800\,\mathrm{m}$.
Apart from $D^z$, all the parameters in the four OI settings do share the same values.
The variables considered are TG, TX and TN, they are further abbreviated as $g$, $x$ and $n$, respectively.
For example, the notation for the analysis of TG at the $j$th grid point for the OI configuration with $D^z=800\,\mathrm{m}$ is $\mathbf{x}^{g,a}_{j} (\boldsymbol{\lambda}_4)$ (see also Tab.~\ref{tab:not}).
In addition to the analysis, the Integral Data Influence \citep[IDI,][]{ULS2008} is available for each point in each field.
IDI is an indicator ranging from 0 to 1 that summarizes the amount of information used in the OI that is derived from nearby observations.
At the $j$th grid point, if we consider for instance TX, the notation $\mathbf{x}^{x,\mathrm{IDI}}_{j} (\boldsymbol{\lambda}_2)$ indicates IDI obtained with $D^z=400\,\mathrm{m}$.
Given a particular set of $\boldsymbol{\lambda}$ values, grid points where the IDI values are close to 0 will have the analysis almost identical to the large-scale pseudo-background, since no information from nearby observations has been used in the analysis.
Then, the spatial support of the analysis is mostly determined by the pseudo-background characteristics.
On the other hand, grid points where the IDI values are close to 1 will use nearby observations to adjust the analysis.
In this case, the spatial support of the analysis is mostly determined by the OI settings and the spatial support of the analysis temperature field is smaller for smaller values of $D^z$.

The merging is done such that more weight is given to the field with smaller spatial support (i.e. $\boldsymbol{\lambda}_1$) and weight has been given to the other (coarser) fields only where they provide additional information.
In this way, we make use of the observational network in an optimal way by providing a higher resolution field where the network is denser.
The relative content of information of a field with respect to another field is evaluated through a function inspired by Shannon's measure of information content, similarly to what reported for the probability density functions in \cite{TARboo}, Sec. 1.2.5.
In our equations, IDI will replace probability density functions.
The fields are ordered from the one with the smallest spatial supports, which is that obtained with $\boldsymbol{\lambda}_1$, to the one with the largest spatial supports, when the OI is performed with $\boldsymbol{\lambda}_4$.
Then, relative content of information in one field with respect to its immediate antecedent field with smaller spatial supports is computed.
The first field, the one with $\boldsymbol{\lambda}_1$, is assigned the complementary of the relative content of information of the second field.
The weights are the normalized relative contents of information.
The procedure is described in mathematical terms in the equations that follows.

The TG analysis at the $j$th grid point is obtained as a weighted average of the four OI analysis:
\begin{equation}
  \mathbf{x}^{g,a}_{j} = \sum_{l=1}^{4} \mathbf{w}^l_j \cdot \mathbf{x}^{g,a}_{j} (\boldsymbol{\lambda}_l)
  \label{eq:merg1}
\end{equation}
where the weights with $l=2,\ldots,4$ are obtained as:
\begin{equation}
  \mathbf{w}^l_j = \boldsymbol{\alpha}^{-1}_j \cdot \mathbf{x}^{g,\mathrm{IDI}}_{j} (\boldsymbol{\lambda}_l) \cdot \mathrm{log} \left( \frac{\mathbf{x}^{g,\mathrm{IDI}}_{j} (\boldsymbol{\lambda}_l) }{\mathbf{x}^{g,\mathrm{IDI}}_{j} (\boldsymbol{\lambda}_{l-1})} \right)
  \label{eq:merg2}
\end{equation}
The weight for $l=1$ is set to:
\begin{equation}
  \mathbf{w}^1 = \boldsymbol{\alpha}^{-1} \cdot (1-\mathbf{w}^2)
  \label{eq:merg3}
\end{equation}
and $\alpha$ is the vector of normalization factors used to enforce that $ \sum_{l=1}^{4} \mathbf{w}^l_j = 1,\; \forall j=1,\ldots,m$.
Note that to avoid problems in Eq.~\eqref{eq:merg2}, all the IDI values smaller than 0.0000001 have been replaced with 0.0000001.

\begin{figure}[h!]
    \includegraphics[width=\textwidth]{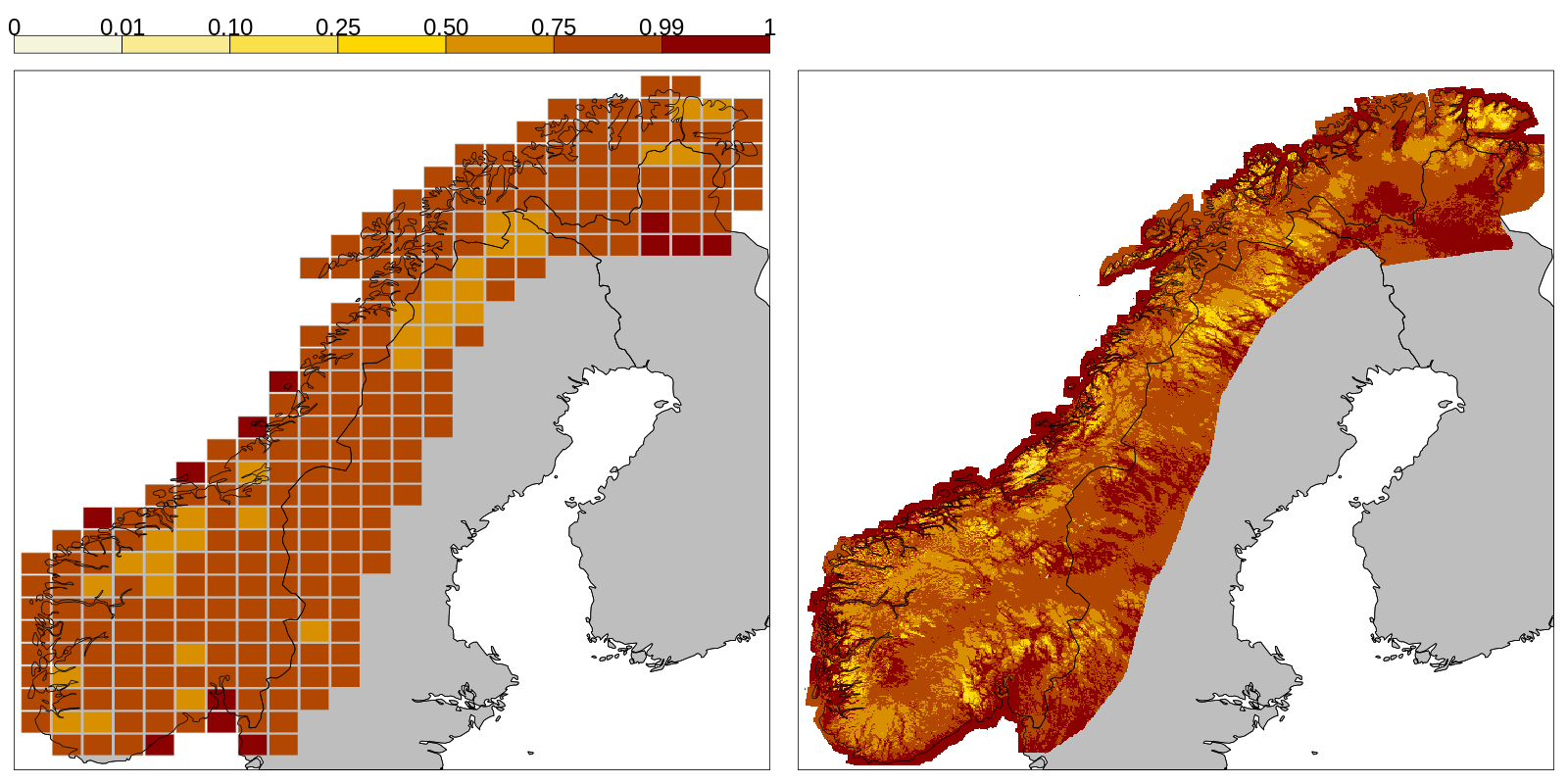}
    \caption{Normalized weights $\mathbf{w}^1$ ($D^z=210\,\mathrm{m}$) for TG, averaged over 2019. The left panel shows the aggregated field over square boxes with sides of 50 km. The right panel shows the original field on the 1 km grid.}
    \label{fig:met_t_w210}
\end{figure}

\begin{figure}[h!]
    \includegraphics[width=\textwidth]{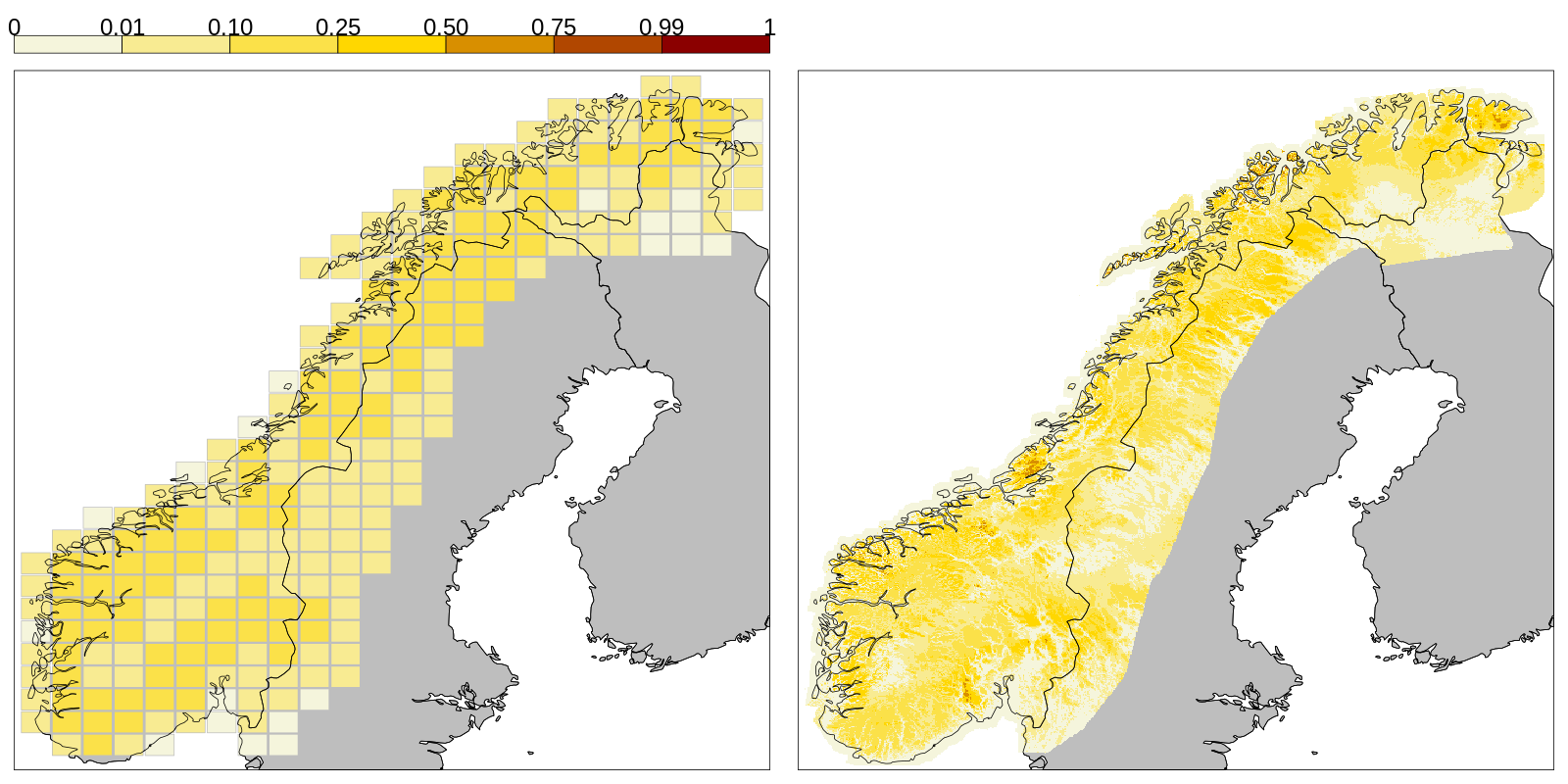}
    \caption{Same as Fig.~\ref{fig:met_t_w210} but for $\mathbf{w}^2$ ($D^z=400\,\mathrm{m}$).}
    \label{fig:met_t_w400}
\end{figure}

\begin{figure}[h!]
    \includegraphics[width=\textwidth]{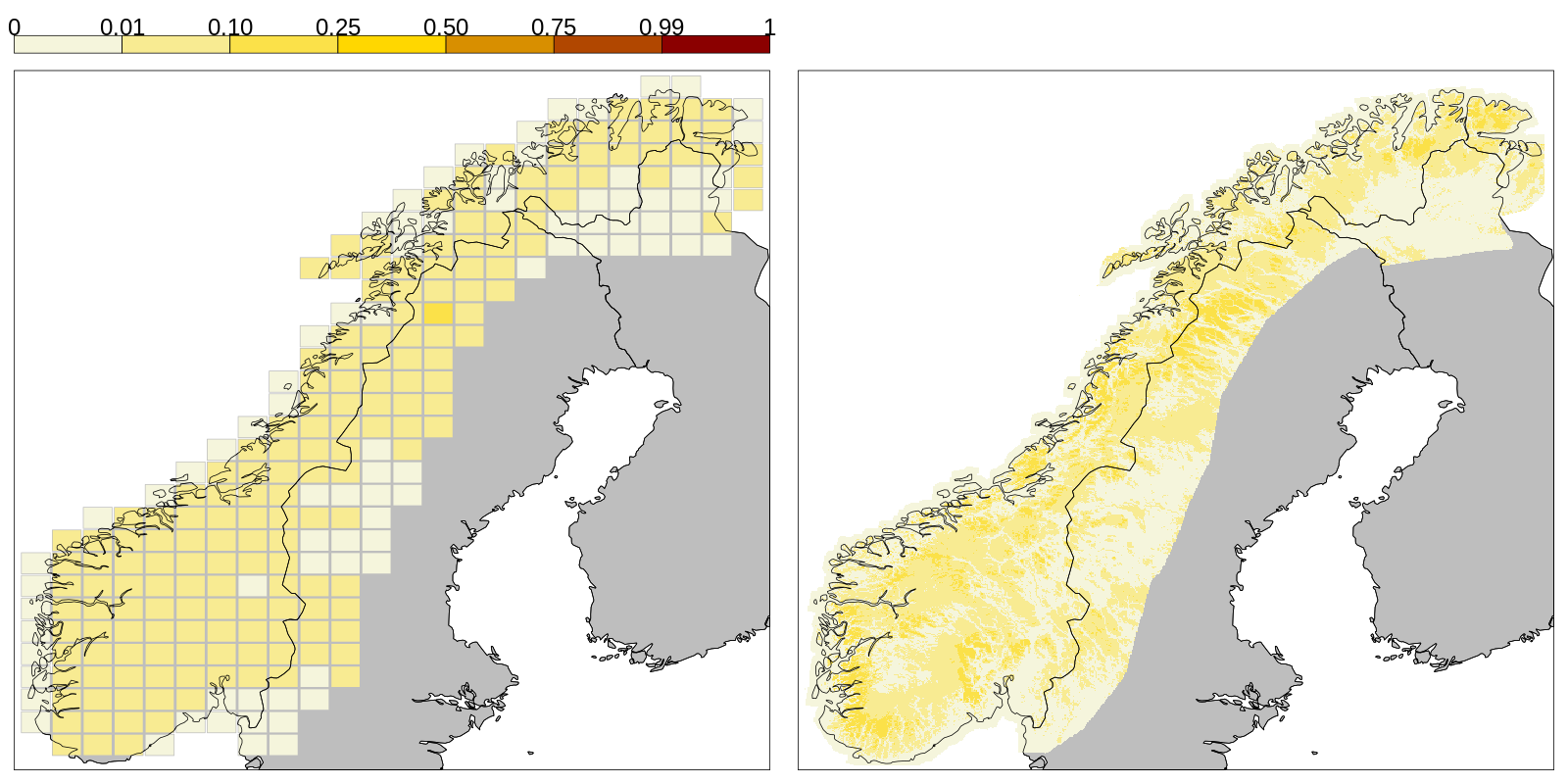}
    \caption{Same as Fig.~\ref{fig:met_t_w210} but for $\mathbf{w}^3$ ($D^z=600\,\mathrm{m}$).}
    \label{fig:met_t_w600}
\end{figure}

\begin{figure}[h!]
    \includegraphics[width=\textwidth]{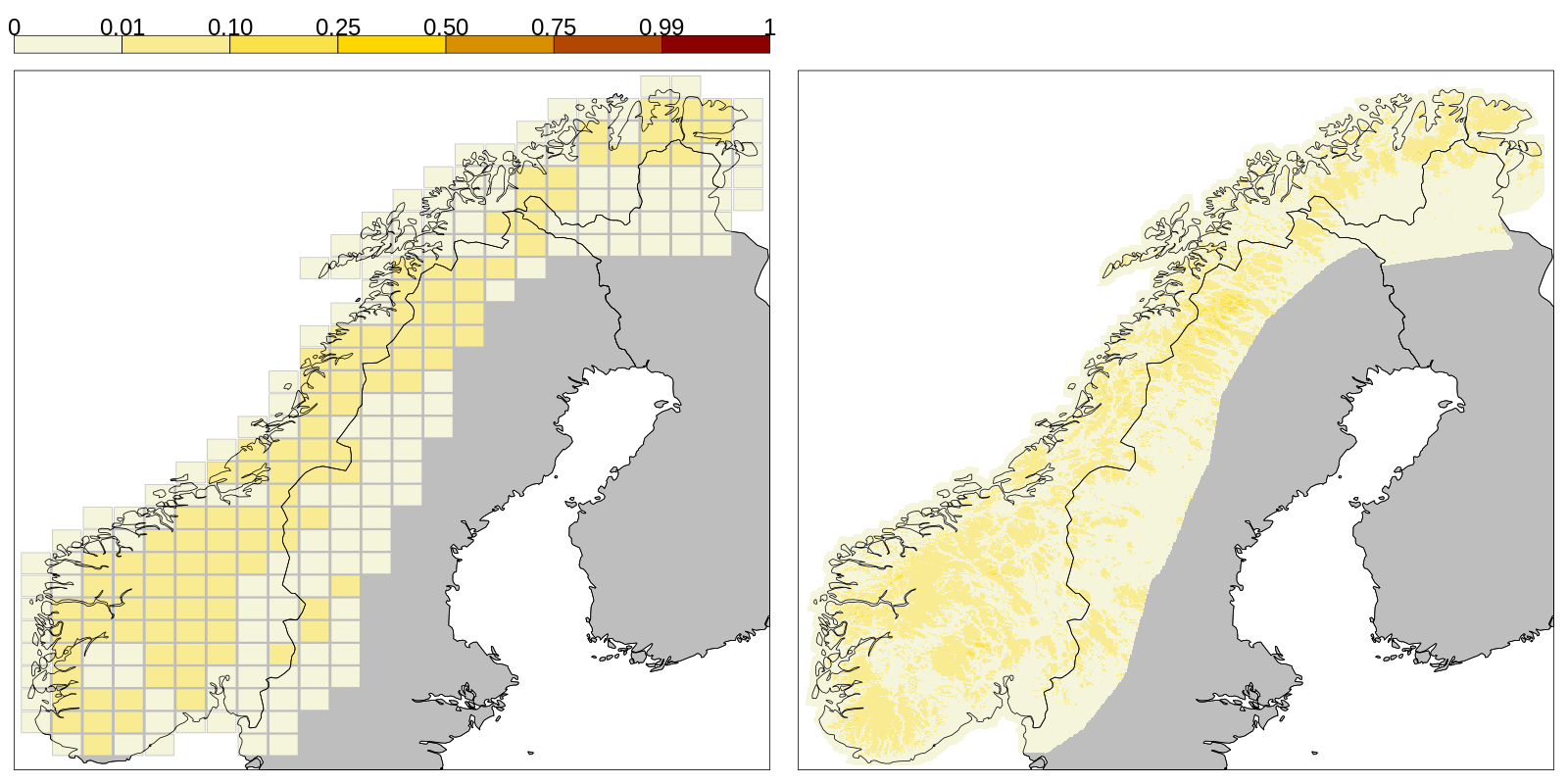}
    \caption{Same as Fig.~\ref{fig:met_t_w210} but for $\mathbf{w}^4$ ($D^z=800\,\mathrm{m}$).}
    \label{fig:met_t_w800}
\end{figure}

In Figs.~\ref{fig:met_t_w210}-~\ref{fig:met_t_w800} examples of the normalized weights of Eqs.~\eqref{eq:merg2}-~\eqref{eq:merg3} are shown.
The figures show the averaged weights over the days within year 2019 for TG.
In the right panels, the fields on the original seNorge\_2018 grid are shown. 
In the left panels, the fields have been aggregated on a coarse resolution grid with square boxes of side length 50 km.
In this way, we focus on averages over regions and not on small-scale details.
The weight $\mathbf{w}^1$ in Fig.~\ref{fig:met_t_w210} is by far the largest one and for a few boxes on the coarser grid its value is equal to 1.
The OI analysis obtained with  $\boldsymbol{\lambda}_1$ should have more weight than the others, especially close to an observation.
In mountainous regions where the observational network is sparse, $\mathbf{w}^1$ reaches its smallest values.
The weight $\mathbf{w}^2$ in Fig.~\ref{fig:met_t_w400} has maximum values in the range 0.25-0.5 exactly in those regions where $\mathbf{w}^1$ has its smallest values.
The other two weights, $\mathbf{w}^2$ and $\mathbf{w}^3$, are shown in Figs.~\ref{fig:met_t_w600}-~\ref{fig:met_t_w800} and the information of the corresponding OIs are used to integrate the higher-resolutions OI only in mountainous regions where the observational network is sparse.
The maximum value of $\mathbf{w}^3$ is 0.25, while the maximum value for $\mathbf{w}^4$ is 0.16.

\clearpage

\section{Results \label{sec:results}}
\noindent A comparison between seNorge\_2018 ver. 20.05 and ver. 18.12 is reported in this section for the four variables RR, TG, TX and TN.
In particular, the comparison is based on the most recent 30-year period common to both datasets, which is the time period 1988-2017.
Note that a complete study to characterize estimation uncertainties based on cross-validation experiments has not been performed yet for ver. 20.05, while it is available in the paper by \cite{lussana2019essd} for ver. 18.12.

\begin{figure}[h!]
    \includegraphics[width=\textwidth]{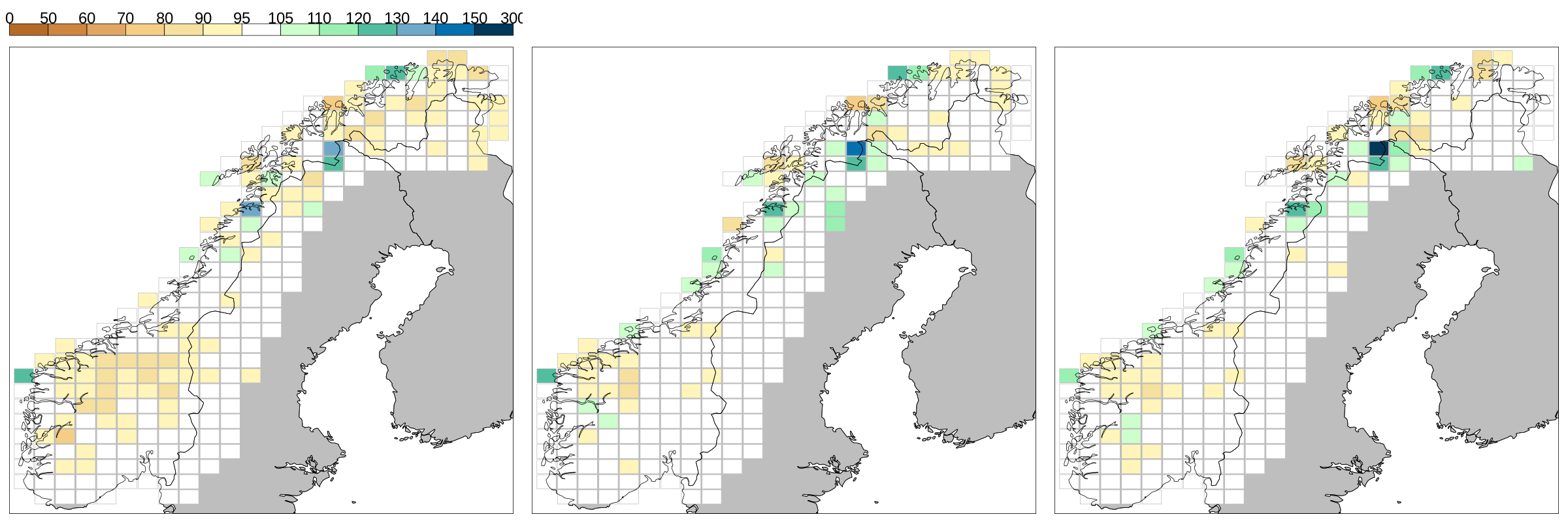}
    \caption{Monthly total precipitation amounts for seNorge\_2018 ver. 20.05, as percentage of ver. 18.12, averaged over the 30-year period from 1988 to 2017 for winter months: December, in the left panel; January, in the middle; February, in the right panel. The fields of averaged values over square boxes with sides of 50 km are shown in the panels.}
    \label{fig:res_prcptot_djf}
\end{figure}

The comparison of total precipitation amounts is shown in Fig.~\ref{fig:res_prcptot_djf} for winter months and in Figs.~\ref{fig:res_prcptot_mam}-~\ref{fig:res_prcptot_son} for the other seasons.
The comparison is performed on monthly precipitation totals averaged over the period 1988-2017.
The results show the amounts of ver. 20.05 as percentages of the amounts of ver. 18.12.
For example, where the monthly total precipitation of ver. 20.05 is approximately equal to the amount of ver. 18-12 ($\pm$5\%), then the corresponding square box will be filled with the white color.
Green/blue colors indicate boxes where ver. 20.05 have larger precipitation amounts than ver. 18.12.
Yellow/brown colors indicate boxes where ver. 20.05 have smaller precipitation amounts than ver. 18.12.
For the majority of the boxes the two versions are rather similar, that is $\pm$5\% or $\pm$10\%.
However, for some boxes in the mountains or along the coast the differences are significant (up to 300\% for two boxes in northern Sweden, close to the boundary with Norway).
The reasons are the different observational dataset used and the variations in the reference field used for the OI.
If compared to the other seasons, the summer months shown in Fig.~\ref{fig:res_prcptot_jja} display a different pattern.
Compared to ver. 18.12, ver. 20.05 has less precipitation along the coast and larger precipitation amounts inland.

\begin{figure}[h!]
    \includegraphics[width=\textwidth]{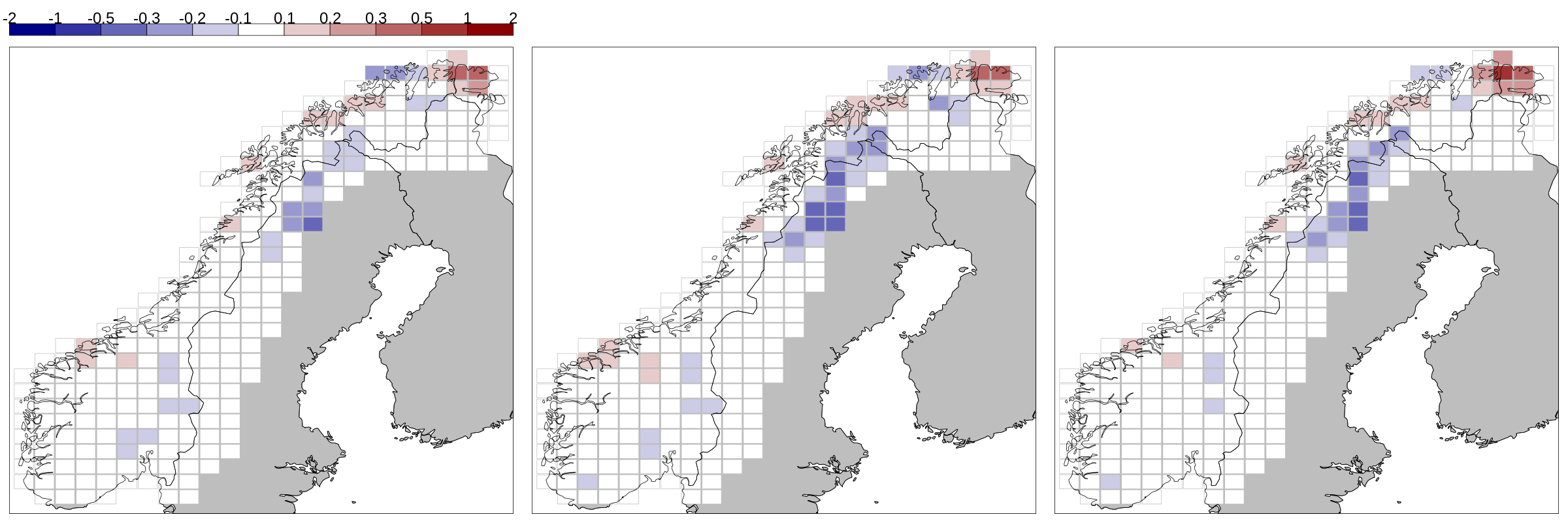}
    \caption{Differences between monthly mean daily temperature of seNorge\_2018 ver. 20.05 and ver. 18.12, as $^\circ \mathrm{C}$, averaged over the 30-year period from 1988 to 2017 for winter months: December, in the left panel; January, in the middle; February, in the right panel. The fields of averaged values over square boxes with sides of 50 km are shown in the panels.}
    \label{fig:res_tmm_djf}
\end{figure}

The comparison of daily mean temperatures is shown in Fig.~\ref{fig:res_tmm_djf} for winter months and in Figs.~\ref{fig:res_tmm_mam}-~\ref{fig:res_tmm_son} for the other seasons.
The quantity considered is the monthly mean temperature averaged over the period 1988-2017 and the elaboration is based on daily mean temperatures. Then, the differences between ver. 20.05 and ver. 18.12 are shown in the figures.
The only significant differences occur during the winter, while for the other season the regional averages are almost everywhere $\pm 0.1^\circ\mathrm{C}$.
Fig.~\ref{fig:res_tmm_djf} shows that during winter: ver. 20.05 is colder than ver. 18.12 on the Scandinavian mountains in the Northern part of Sweden; ver. 20.05 is warmer than ver. 18.12 on the northernmost part of Norway, in the region of East-Finnmark.

\begin{figure}[h!]
    \includegraphics[width=\textwidth]{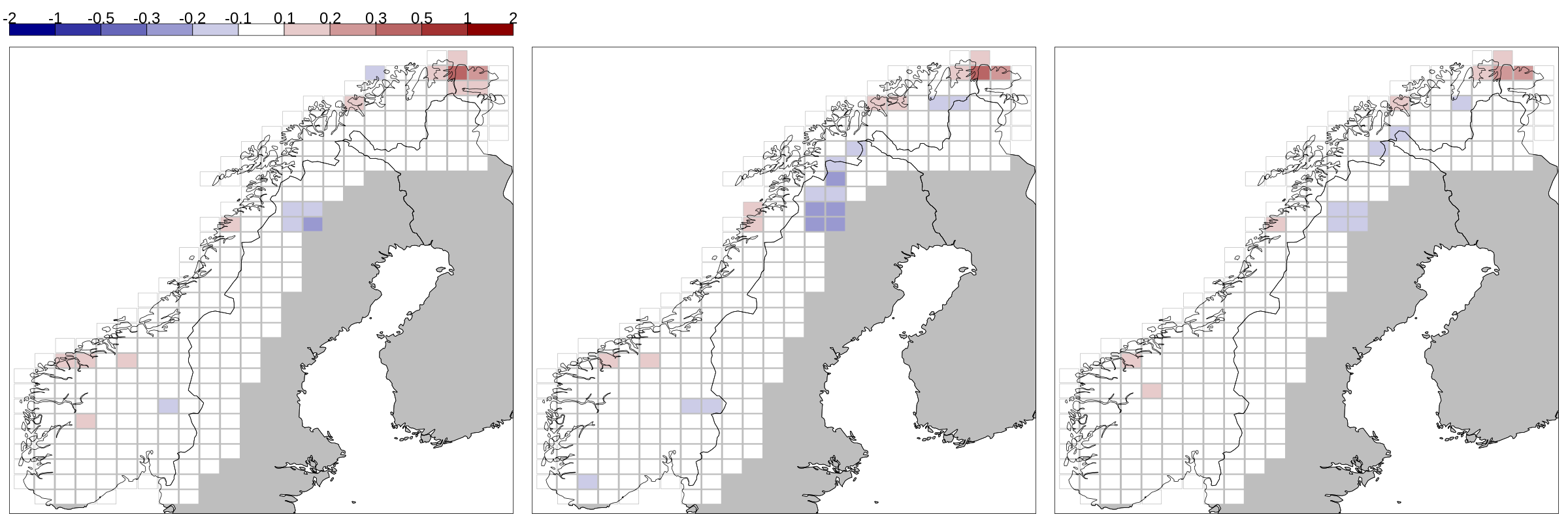}
    \caption{Differences between monthly average of maximum daily temperature of seNorge\_2018 ver. 20.05 and ver. 18.12, as $^\circ \mathrm{C}$, averaged over the 30-year period from 1988 to 2017 for winter months: December, in the left panel; January, in the middle; February, in the right panel. The fields of averaged values over square boxes with sides of 50 km are shown in the panels.}
    \label{fig:res_tmx_djf}
\end{figure}

The comparison of the daily maximum temperature is shown in Fig.~\ref{fig:res_tmx_djf} for winter months and Figs.~\ref{fig:res_tmx_mam}-~\ref{fig:res_tmx_son} for the other months.
The quantity considered is the monthly mean of daily maximum temperatures. The monthly mean is based on the period 1988-2017. Then, the differences between ver. 20.05 and ver. 18.12 are shown in the figures.
In general, the two seNorge\_2018 versions are rather similar. The situation is similar to that of daily mean temperature, but the differences between the two versions are even less pronounced.

\begin{figure}[h!]
    \includegraphics[width=\textwidth]{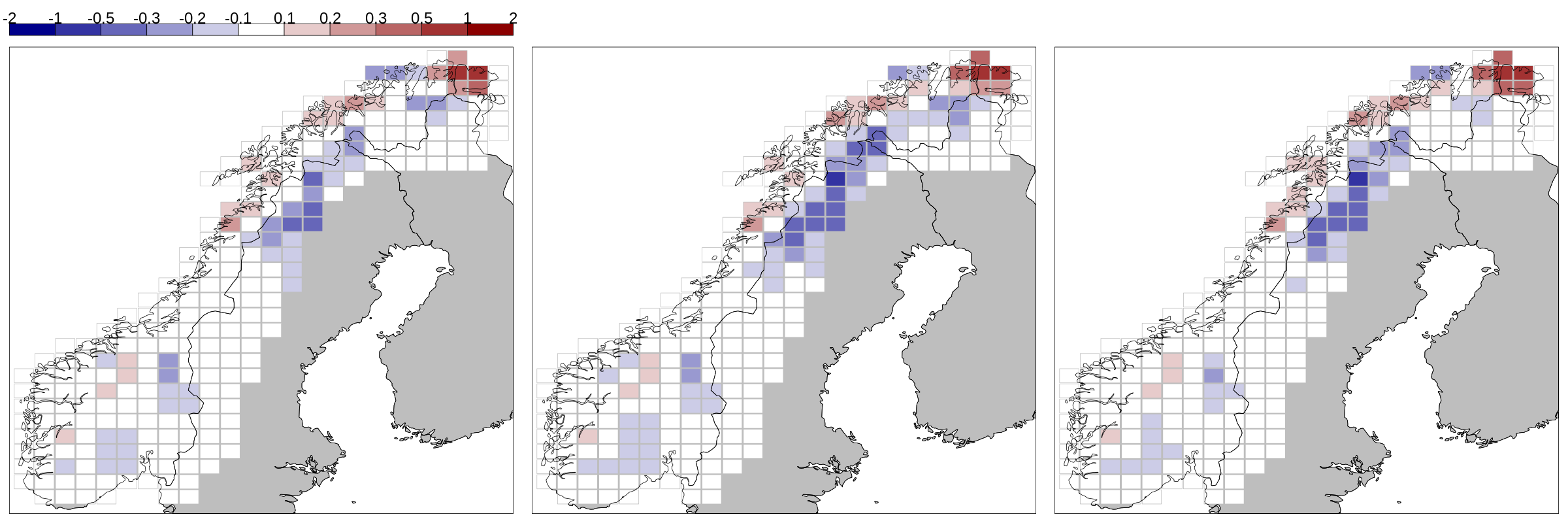}
    \caption{Differences between monthly average of maximum daily temperature of seNorge\_2018 ver. 20.05 and ver. 18.12, as $^\circ \mathrm{C}$, averaged over the 30-year period from 1988 to 2017 for winter months: December, in the left panel; January, in the middle; February, in the right panel. The fields of averaged values over square boxes with sides of 50 km are shown in the panels.}
    \label{fig:res_tmn_djf}
\end{figure}

\begin{figure}[h!]
    \includegraphics[width=\textwidth]{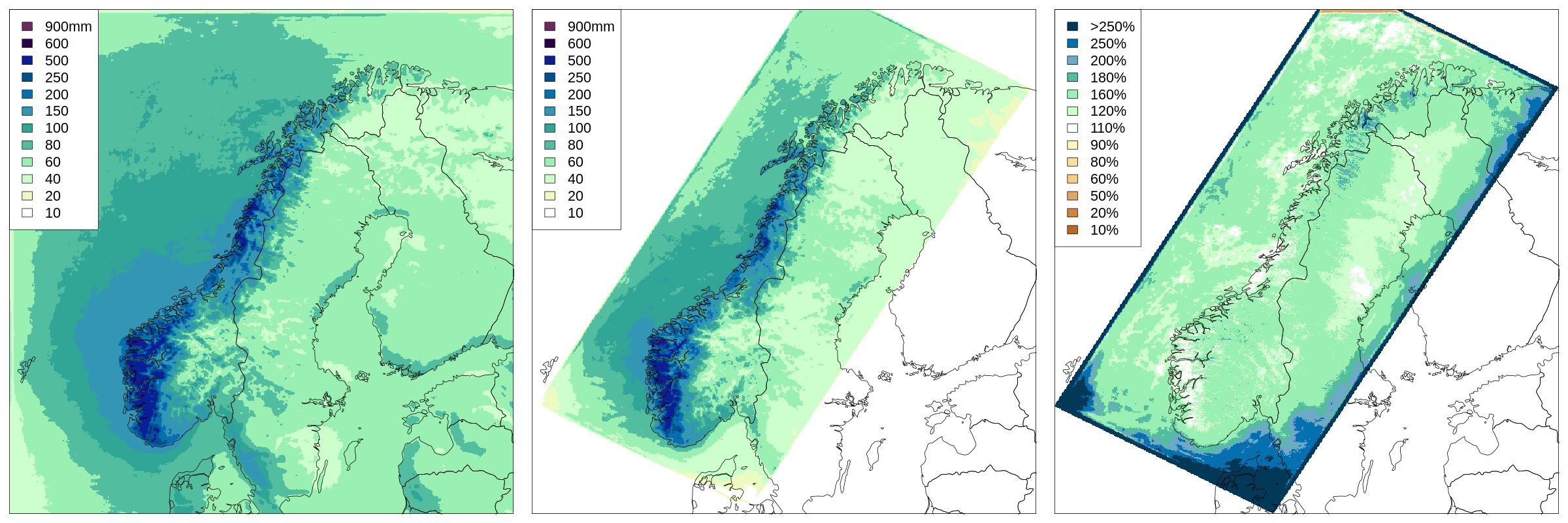}
    \caption{same as Fig.~\ref{fig:rr_ref_comp_year} but for the typical January.}
    \label{fig:rr_ref_comp_month01}
\end{figure}

\begin{figure}[h!]
    \includegraphics[width=\textwidth]{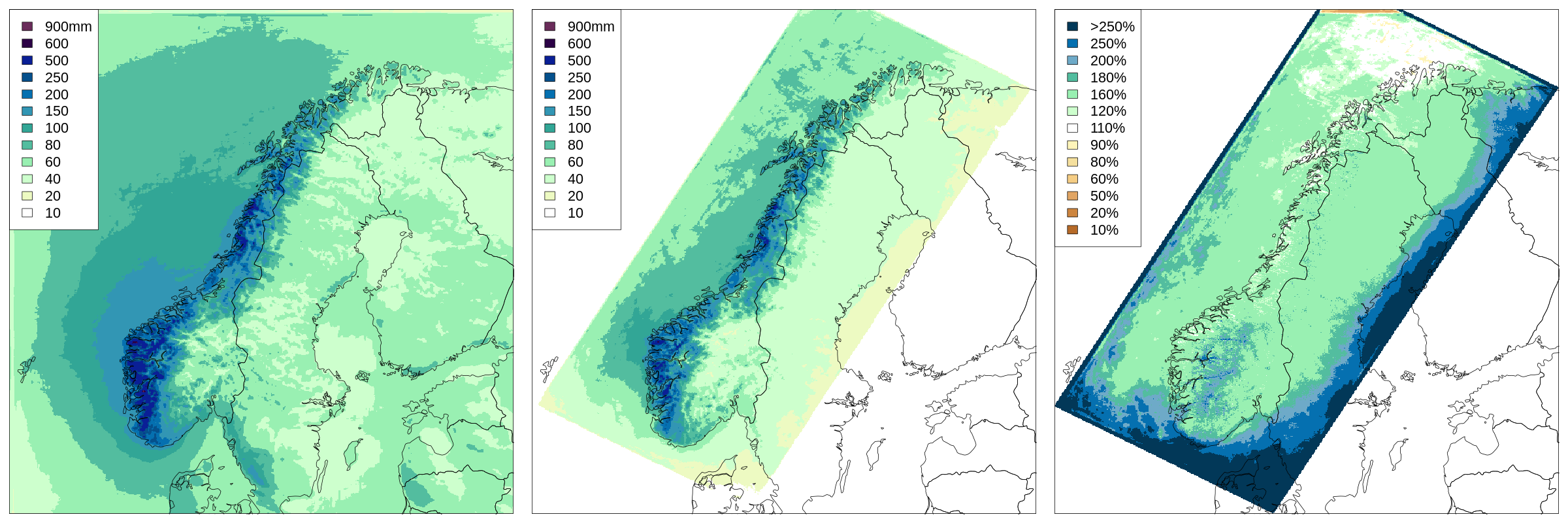}
    \caption{same as Fig.~\ref{fig:rr_ref_comp_year} but for the typical February.}
    \label{fig:rr_ref_comp_month02}
\end{figure}

\begin{figure}[h!]
    \includegraphics[width=\textwidth]{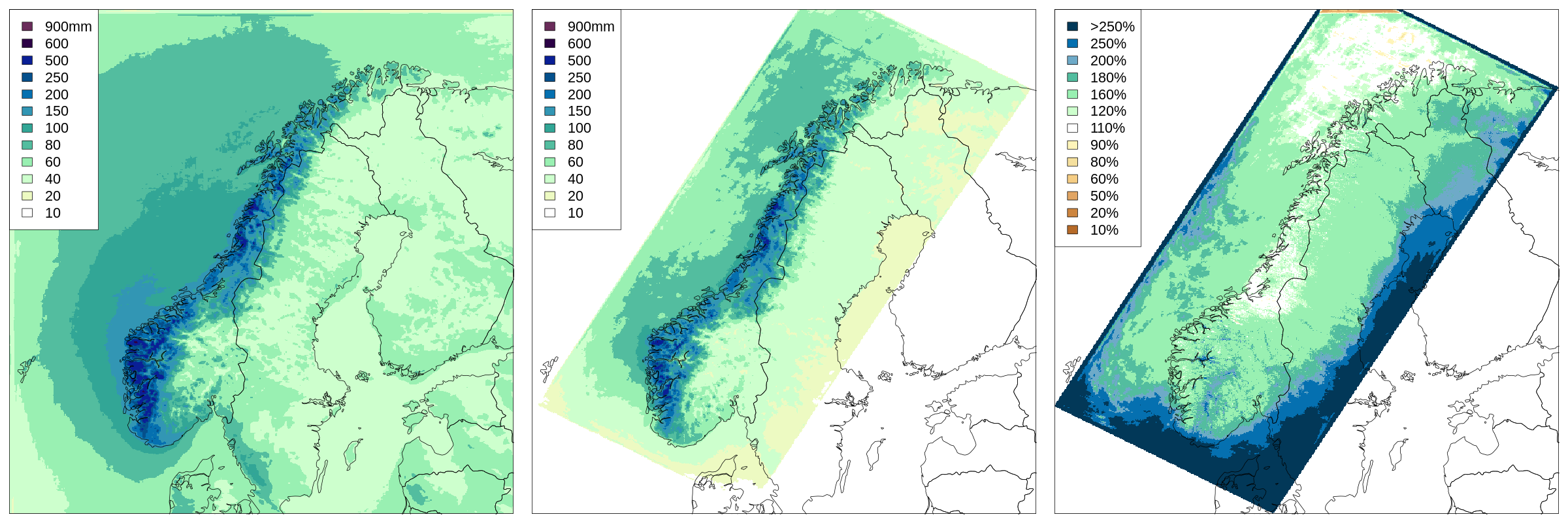}
    \caption{same as Fig.~\ref{fig:rr_ref_comp_year} but for the typical March.}
    \label{fig:rr_ref_comp_month03}
\end{figure}

\begin{figure}[h!]
    \includegraphics[width=\textwidth]{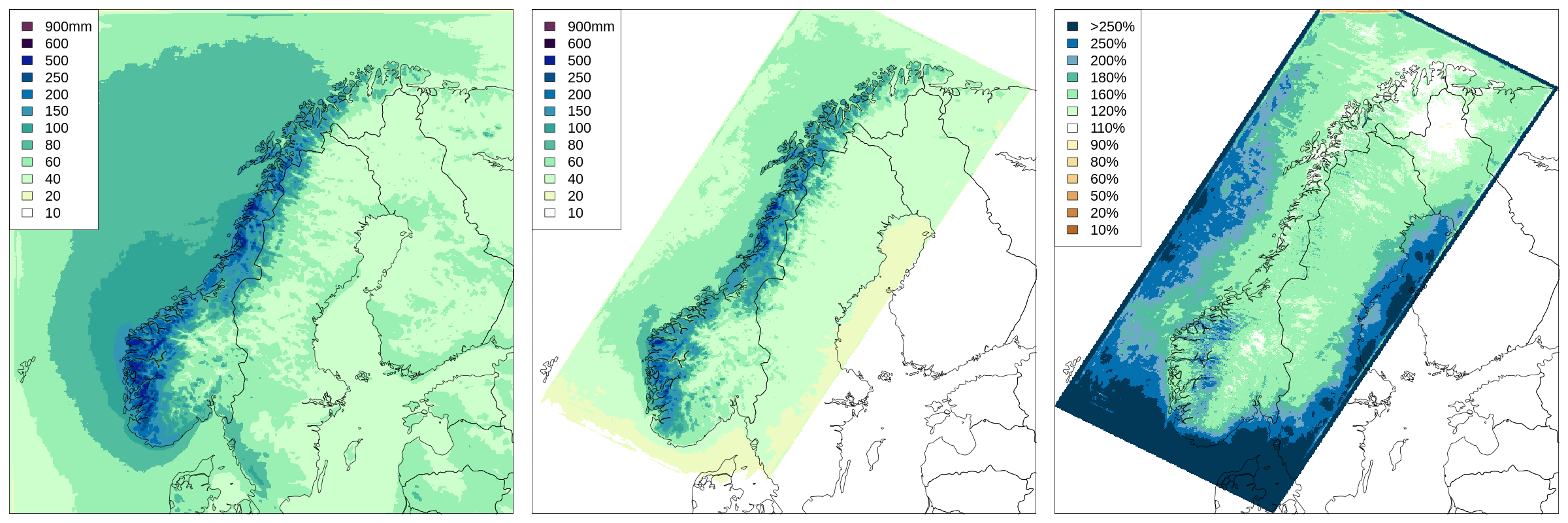}
    \caption{same as Fig.~\ref{fig:rr_ref_comp_year} but for the typical April.}
    \label{fig:rr_ref_comp_month04}
\end{figure}

\begin{figure}[h!]
    \includegraphics[width=\textwidth]{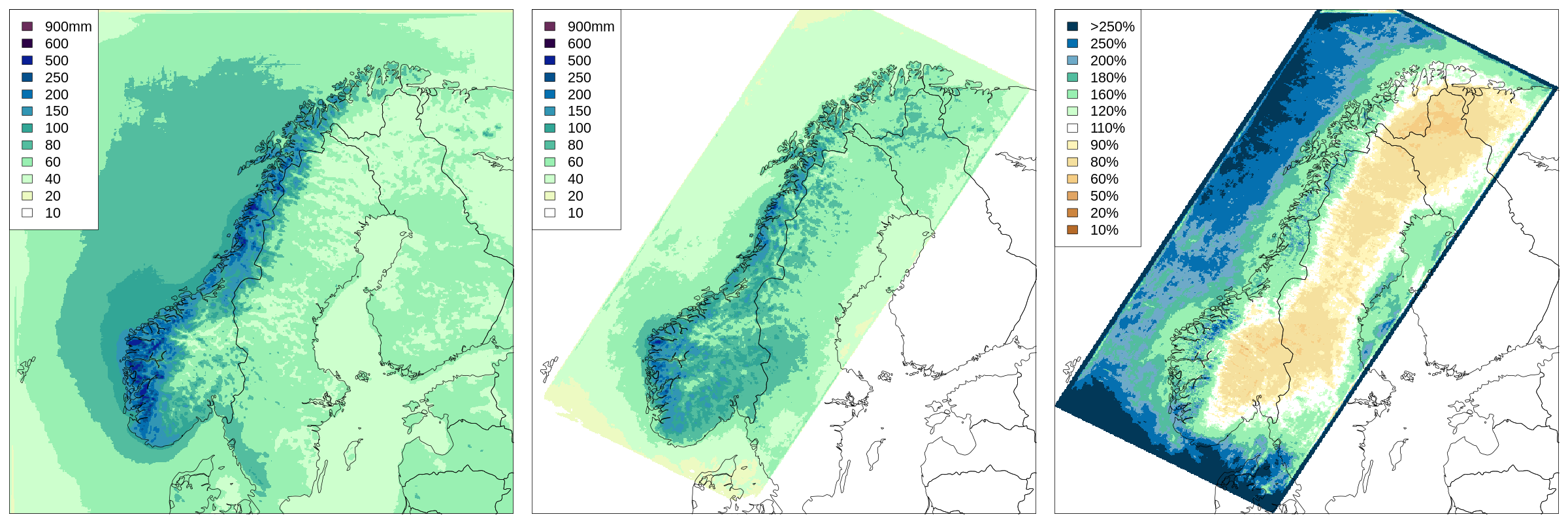}
    \caption{same as Fig.~\ref{fig:rr_ref_comp_year} but for the typical May.}
    \label{fig:rr_ref_comp_month05}
\end{figure}

\begin{figure}[h!]
    \includegraphics[width=\textwidth]{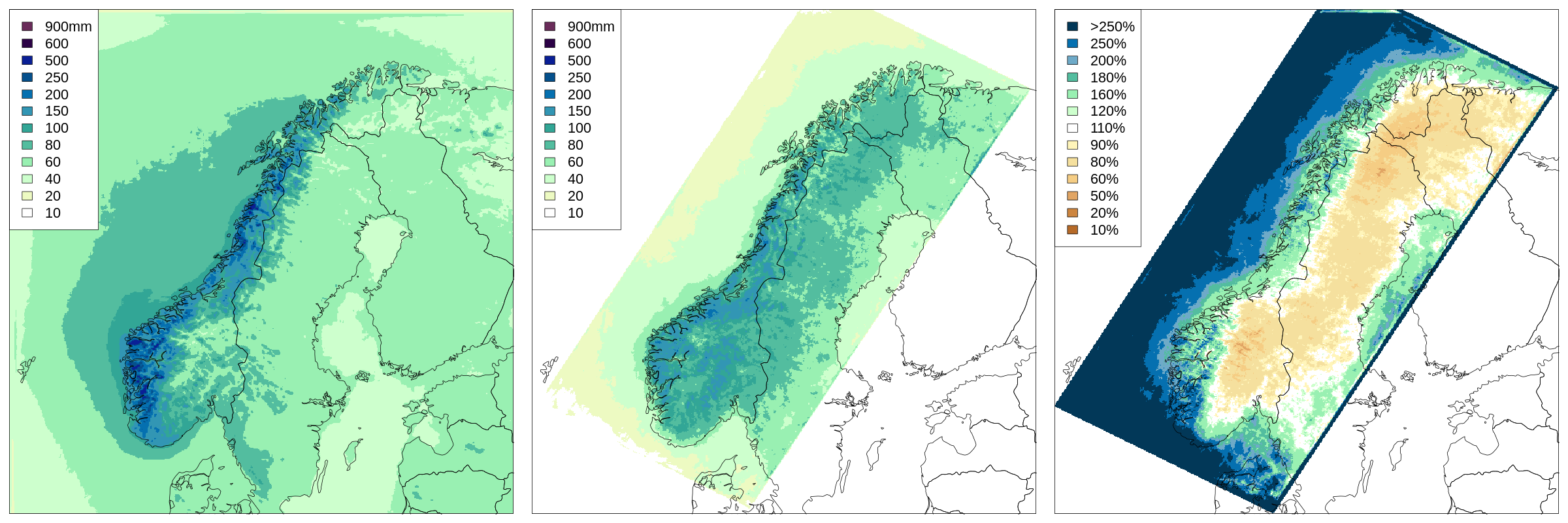}
    \caption{same as Fig.~\ref{fig:rr_ref_comp_year} but for the typical June.}
    \label{fig:rr_ref_comp_month06}
\end{figure}

\begin{figure}[h!]
    \includegraphics[width=\textwidth]{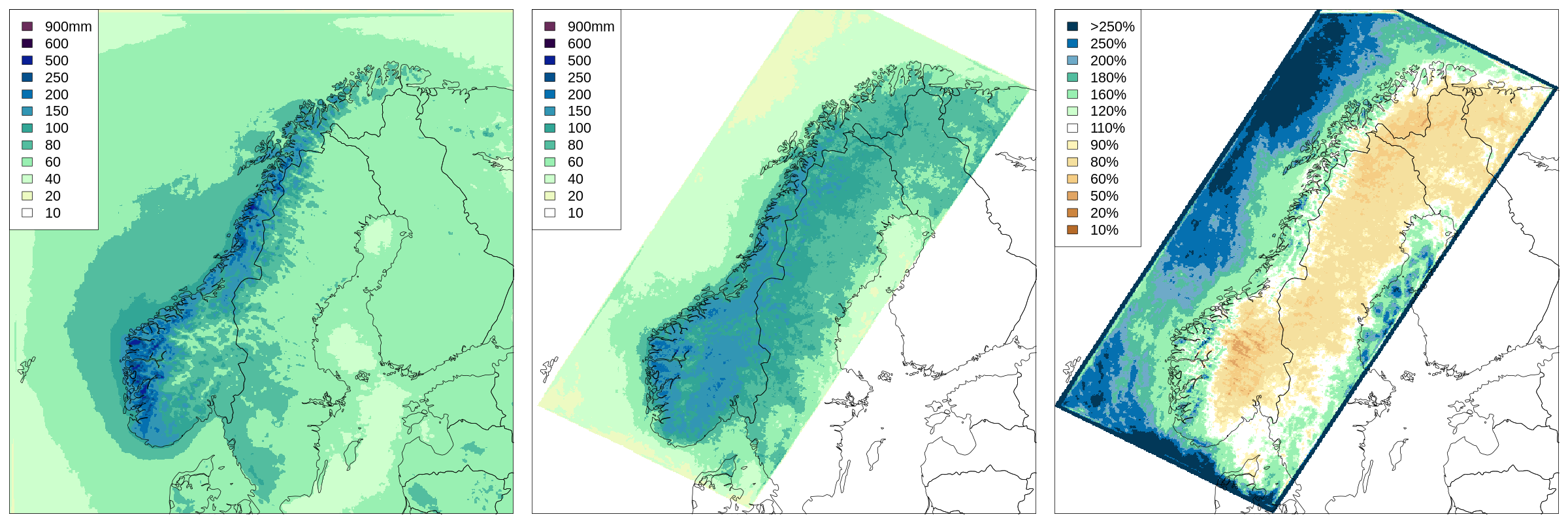}
    \caption{same as Fig.~\ref{fig:rr_ref_comp_year} but for the typical July.}
    \label{fig:rr_ref_comp_month07}
\end{figure}

\begin{figure}[h!]
    \includegraphics[width=\textwidth]{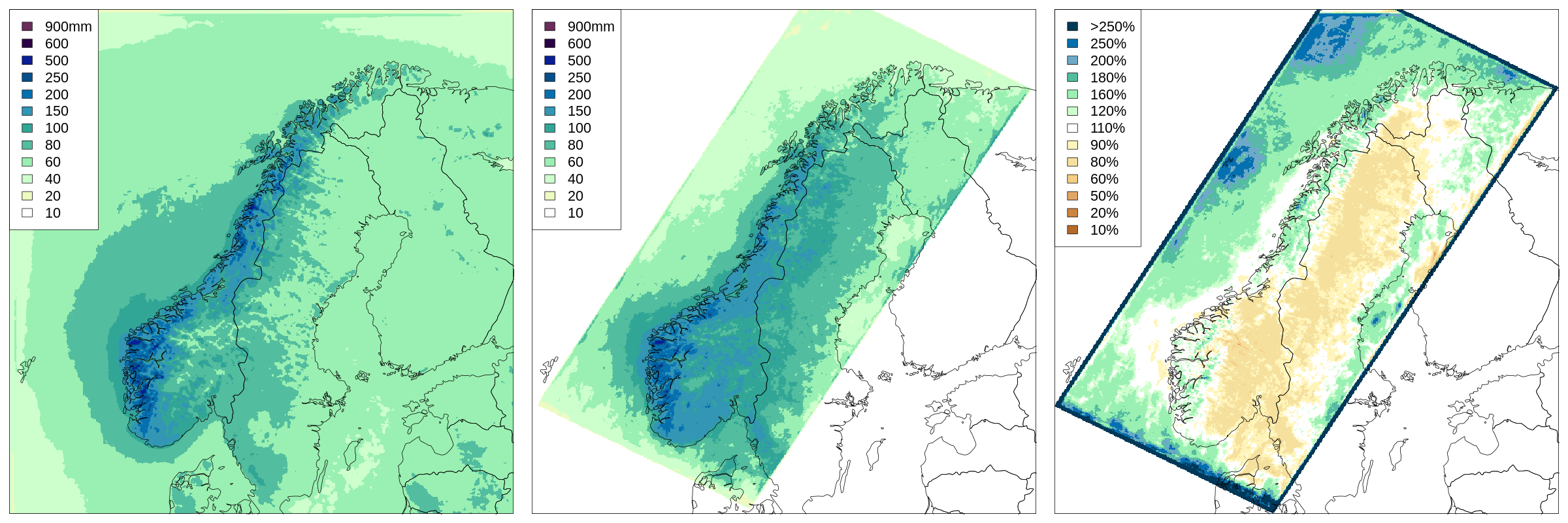}
    \caption{same as Fig.~\ref{fig:rr_ref_comp_year} but for the typical August.}
    \label{fig:rr_ref_comp_month08}
\end{figure}

\begin{figure}[h!]
    \includegraphics[width=\textwidth]{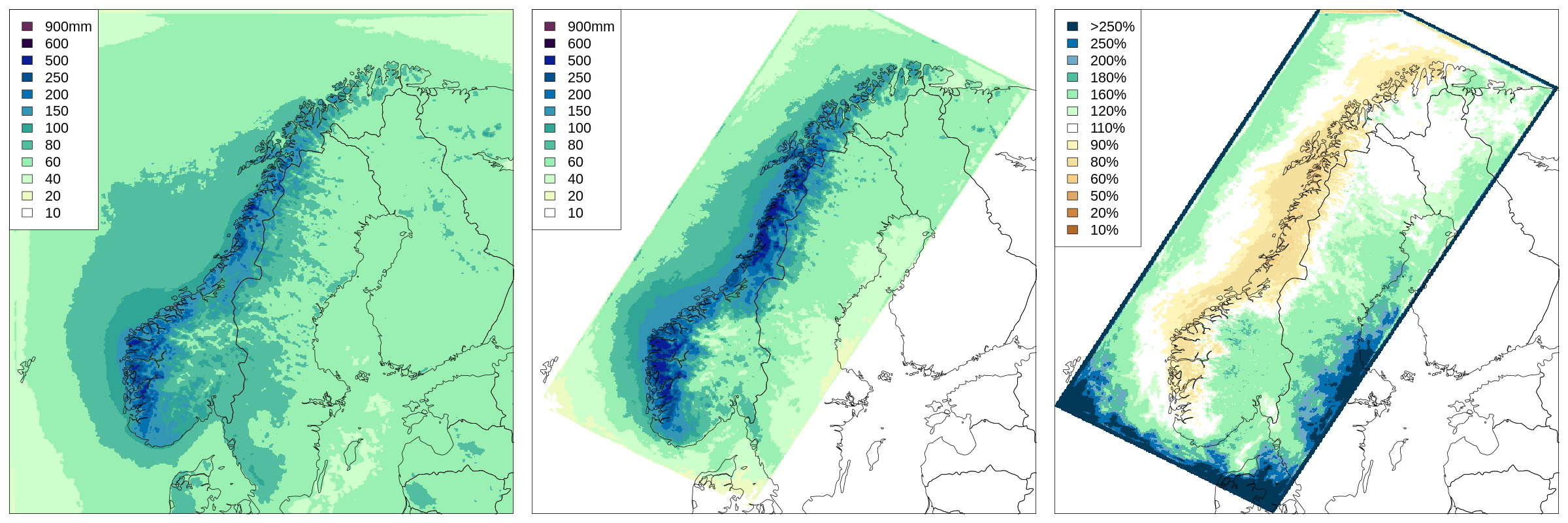}
    \caption{same as Fig.~\ref{fig:rr_ref_comp_year} but for the typical September.}
    \label{fig:rr_ref_comp_month09}
\end{figure}

\begin{figure}[h!]
    \includegraphics[width=\textwidth]{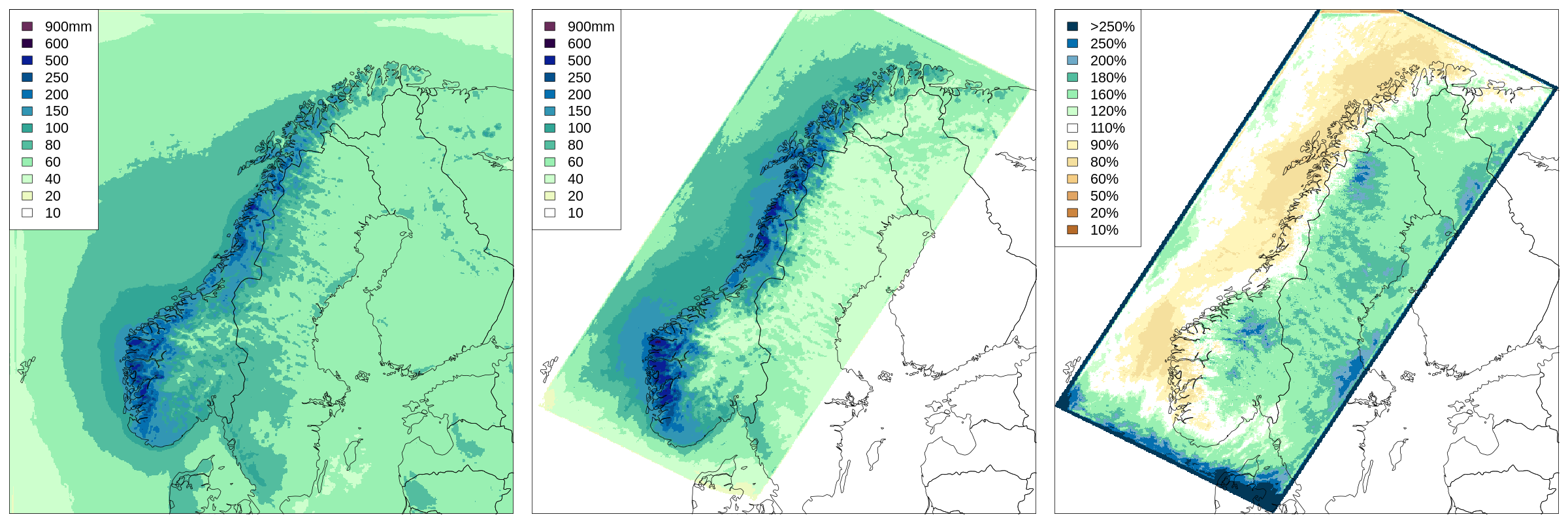}
    \caption{same as Fig.~\ref{fig:rr_ref_comp_year} but for the typical October.}
    \label{fig:rr_ref_comp_month10}
\end{figure}

\begin{figure}[h!]
    \includegraphics[width=\textwidth]{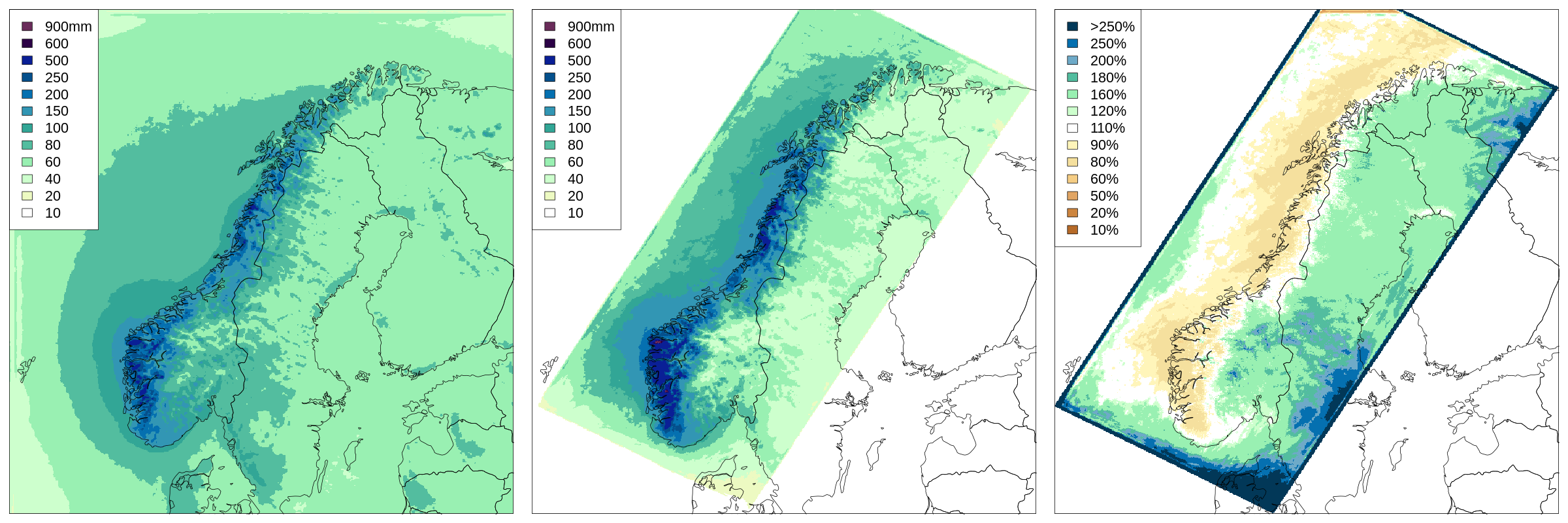}
    \caption{same as Fig.~\ref{fig:rr_ref_comp_year} but for the typical November.}
    \label{fig:rr_ref_comp_month11}
\end{figure}

\begin{figure}[h!]
    \includegraphics[width=\textwidth]{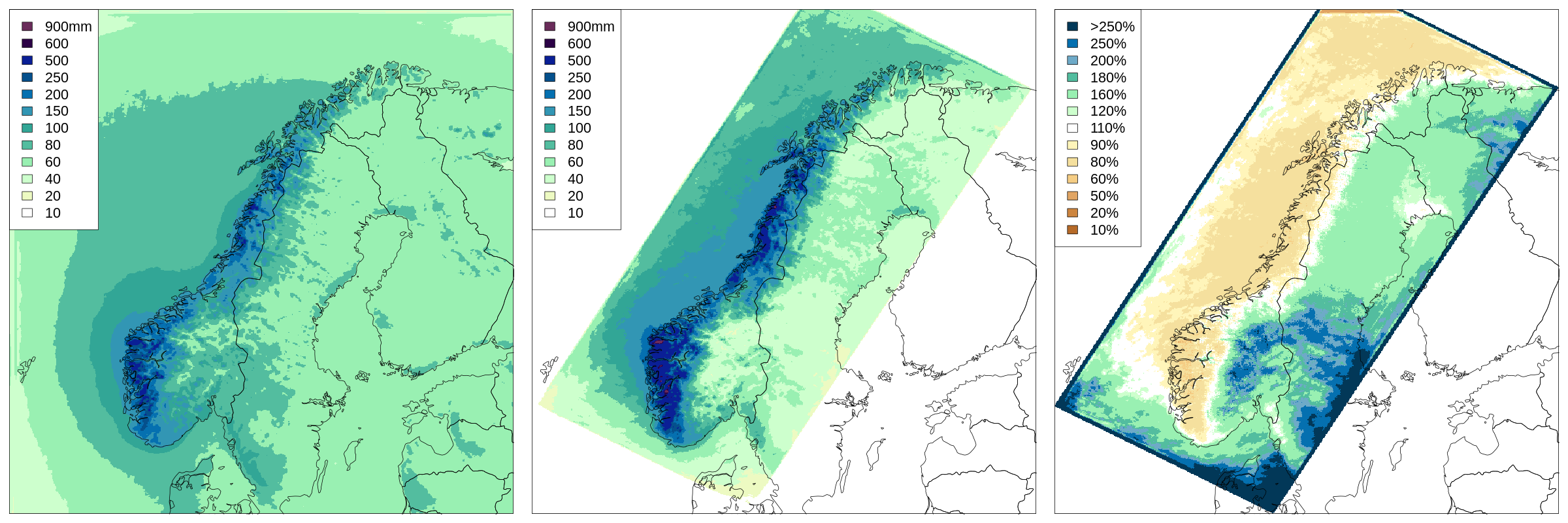}
    \caption{same as Fig.~\ref{fig:rr_ref_comp_year} but for the typical December.}
    \label{fig:rr_ref_comp_month12}
\end{figure}

\begin{figure}[h!]
    \includegraphics[width=\textwidth]{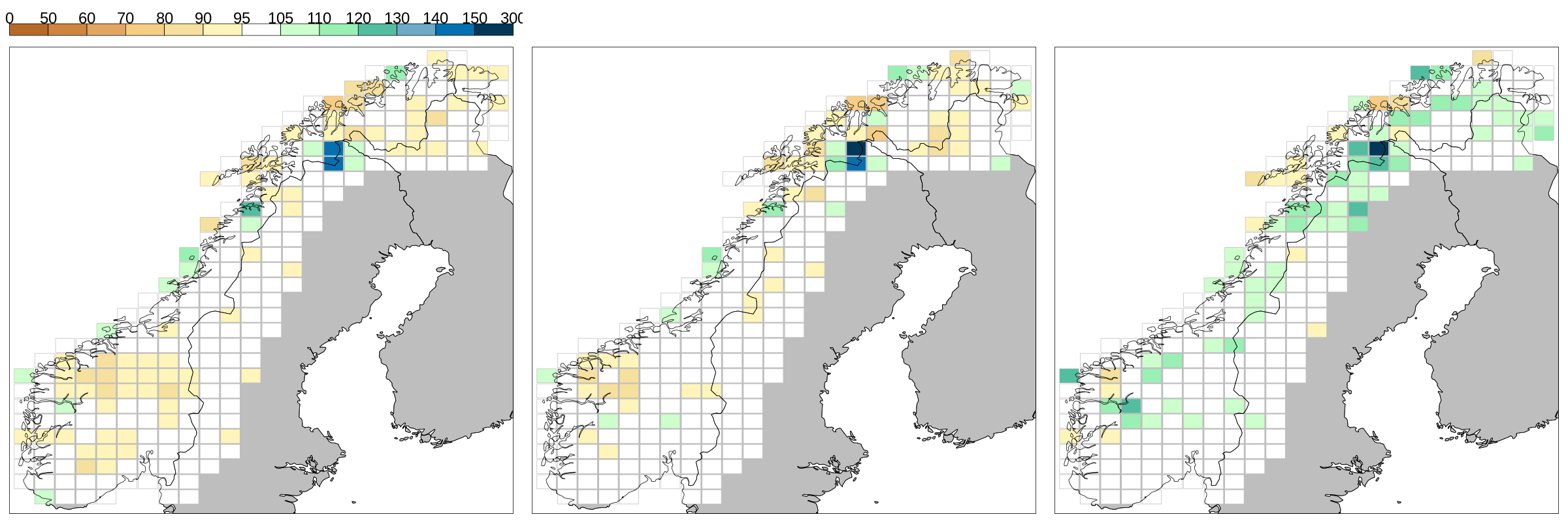}
    \caption{Same as Fig.~\ref{fig:res_prcptot_djf} but for spring months: March, in the left panel; April, in the middle; May, in the right panel.}
    \label{fig:res_prcptot_mam}
\end{figure}

\begin{figure}[h!]
    \includegraphics[width=\textwidth]{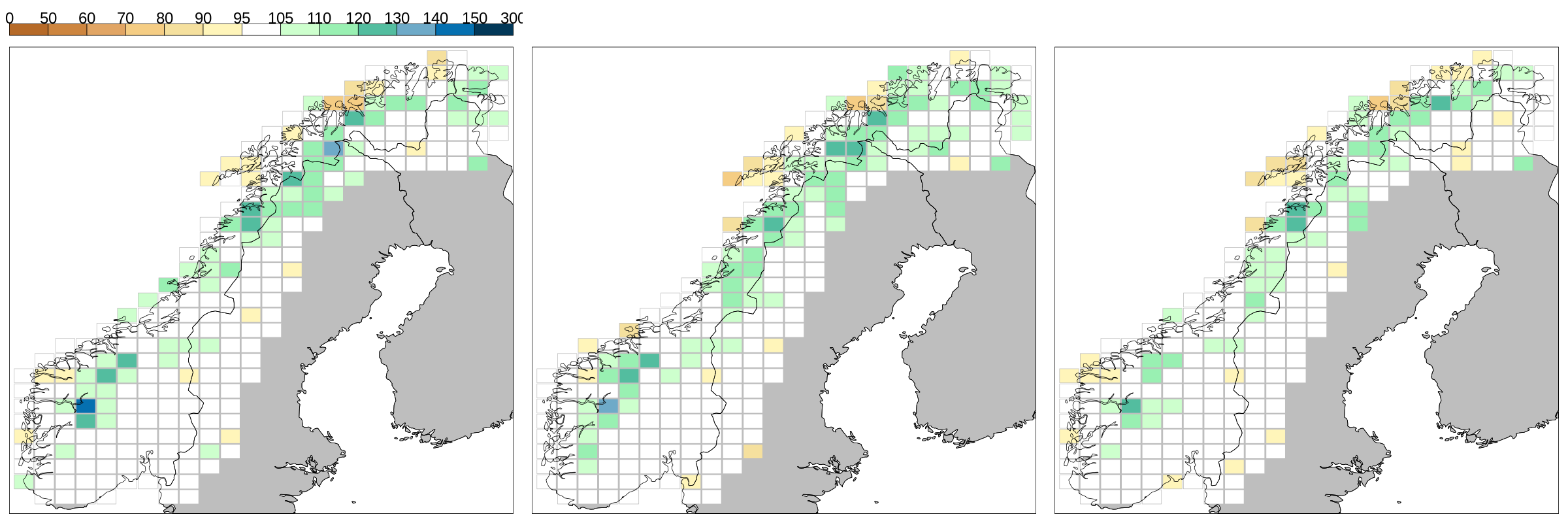}
    \caption{Same as Fig.~\ref{fig:res_prcptot_djf} but for summer months: June, in the left panel; July, in the middle; August, in the right panel.}
    \label{fig:res_prcptot_jja}
\end{figure}

\begin{figure}[h!]
    \includegraphics[width=\textwidth]{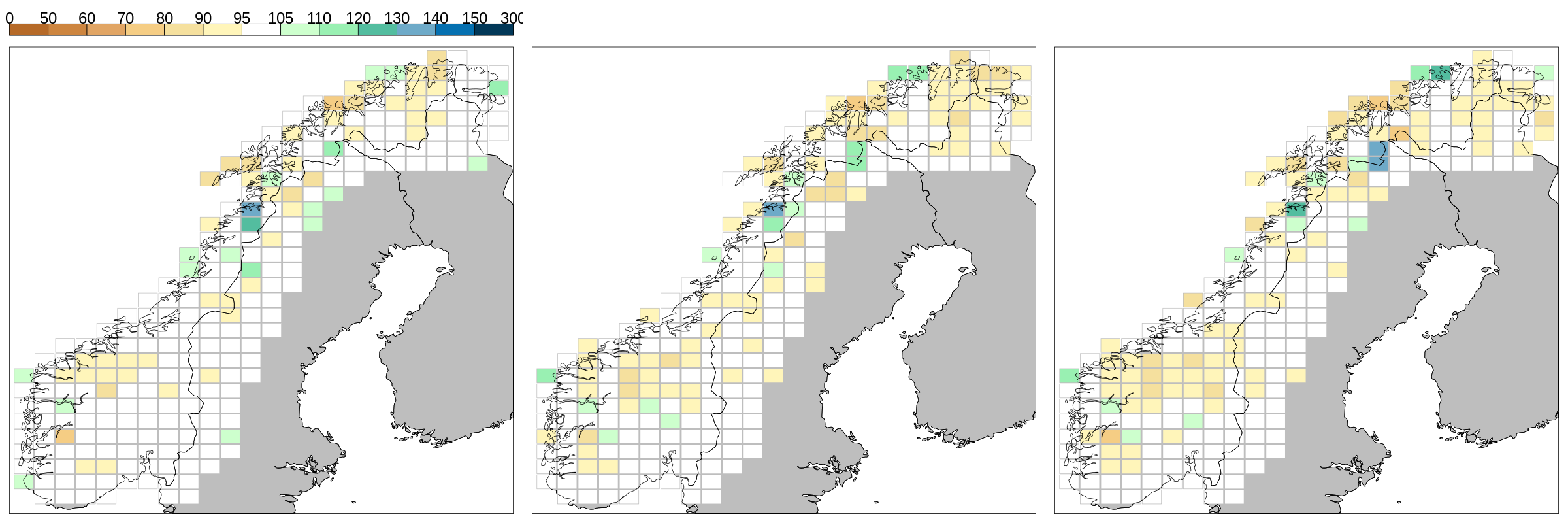}
    \caption{Same as Fig.~\ref{fig:res_prcptot_djf} but for autumn months: September, in the left panel; October, in the middle; November, in the right panel.}
    \label{fig:res_prcptot_son}
\end{figure}

\begin{figure}[h!]
    \includegraphics[width=\textwidth]{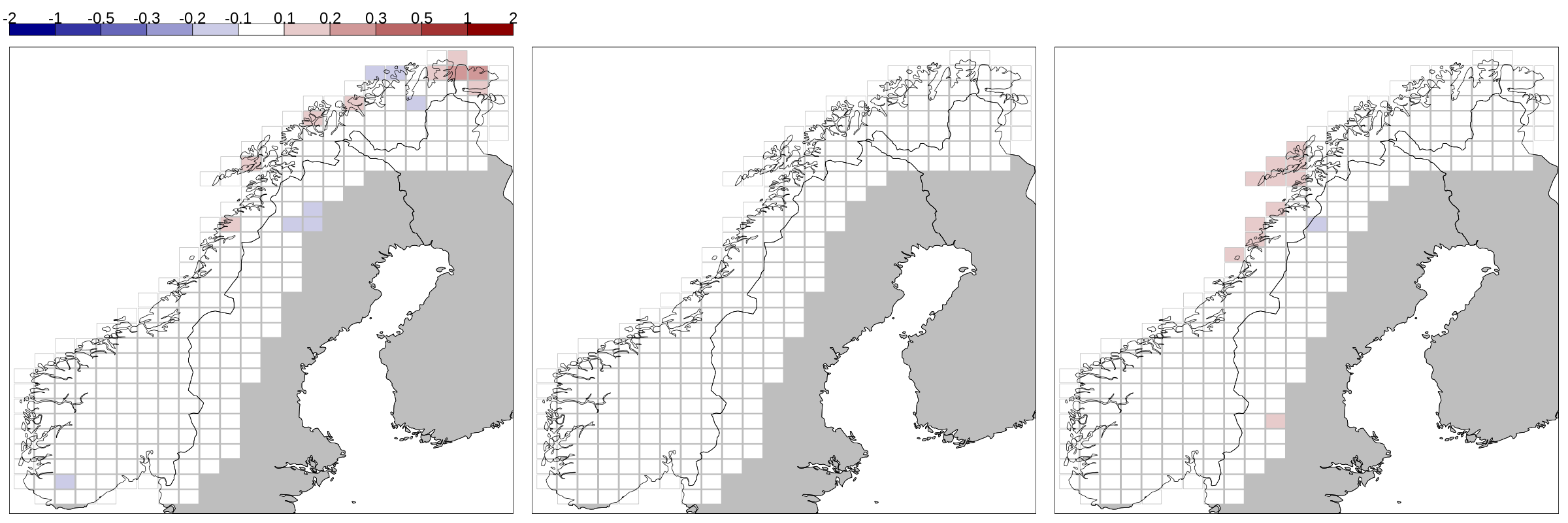}
    \caption{Same as Fig.~\ref{fig:res_tmm_djf} but for spring months: March, in the left panel; April, in the middle; May, in the right panel.}
    \label{fig:res_tmm_mam}
\end{figure}

\begin{figure}[h!]
    \includegraphics[width=\textwidth]{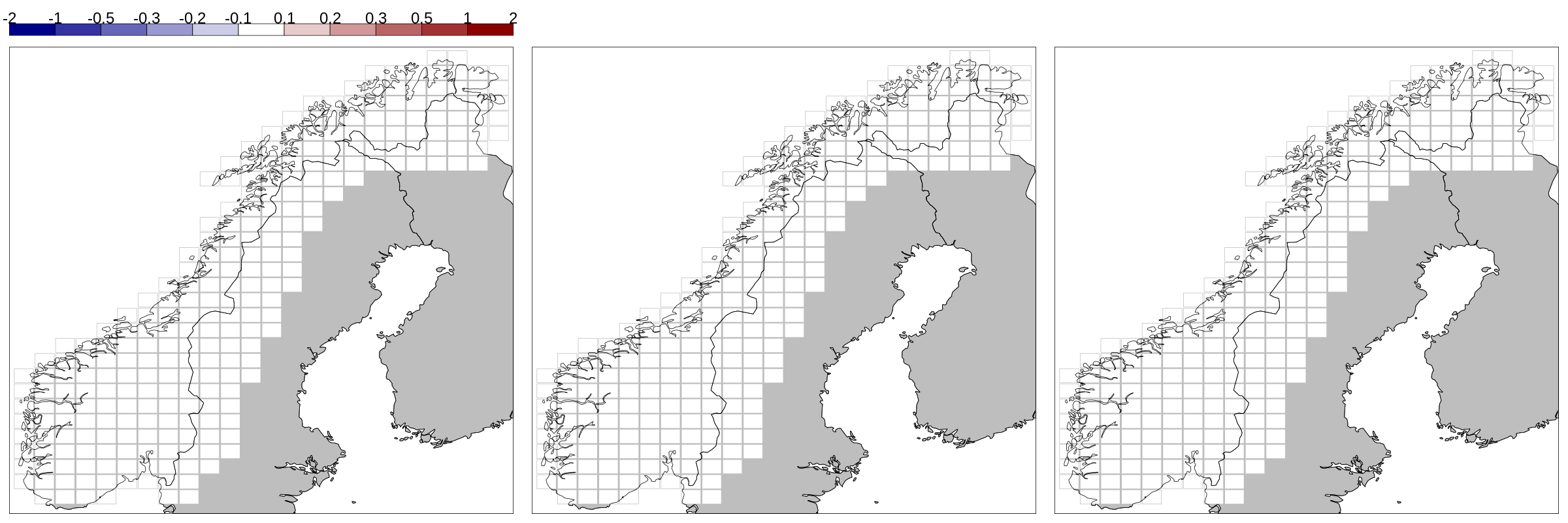}
    \caption{Same as Fig.~\ref{fig:res_tmm_djf} but for summer months: June, in the left panel; July, in the middle; August, in the right panel.}
    \label{fig:res_tmm_jja}
\end{figure}

\begin{figure}[h!]
    \includegraphics[width=\textwidth]{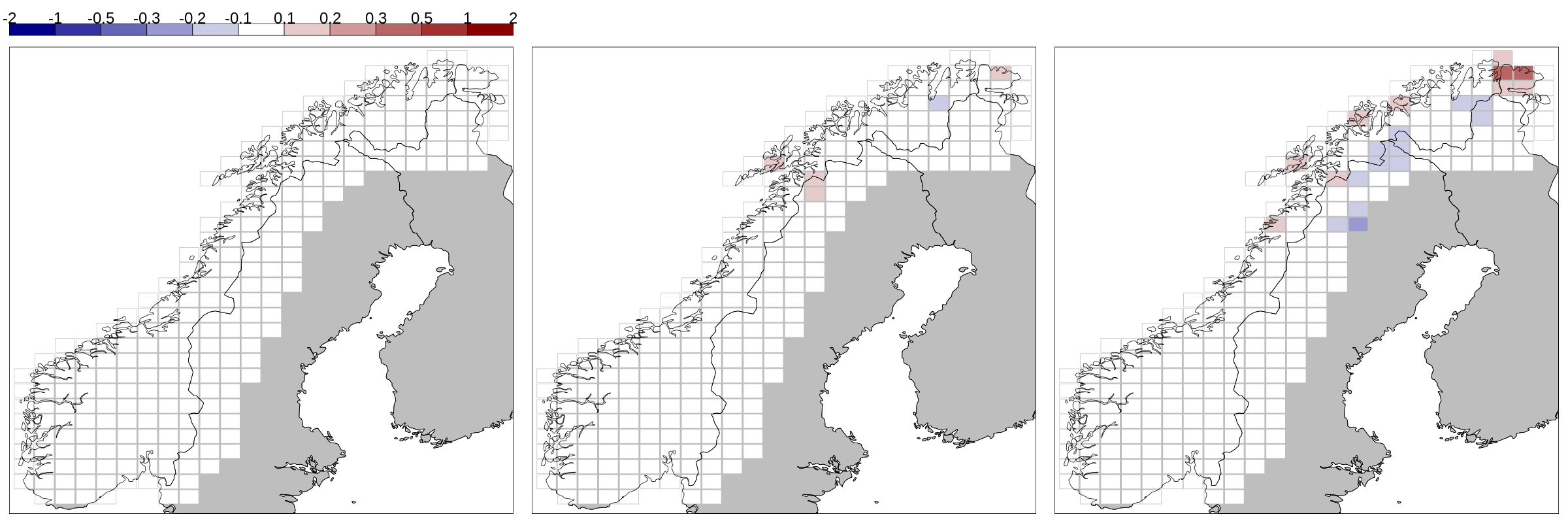}
    \caption{Same as Fig.~\ref{fig:res_tmm_djf} but for autumn months: September, in the left panel; October, in the middle; November, in the right panel.}
    \label{fig:res_tmm_son}
\end{figure}

\begin{figure}[h!]
    \includegraphics[width=\textwidth]{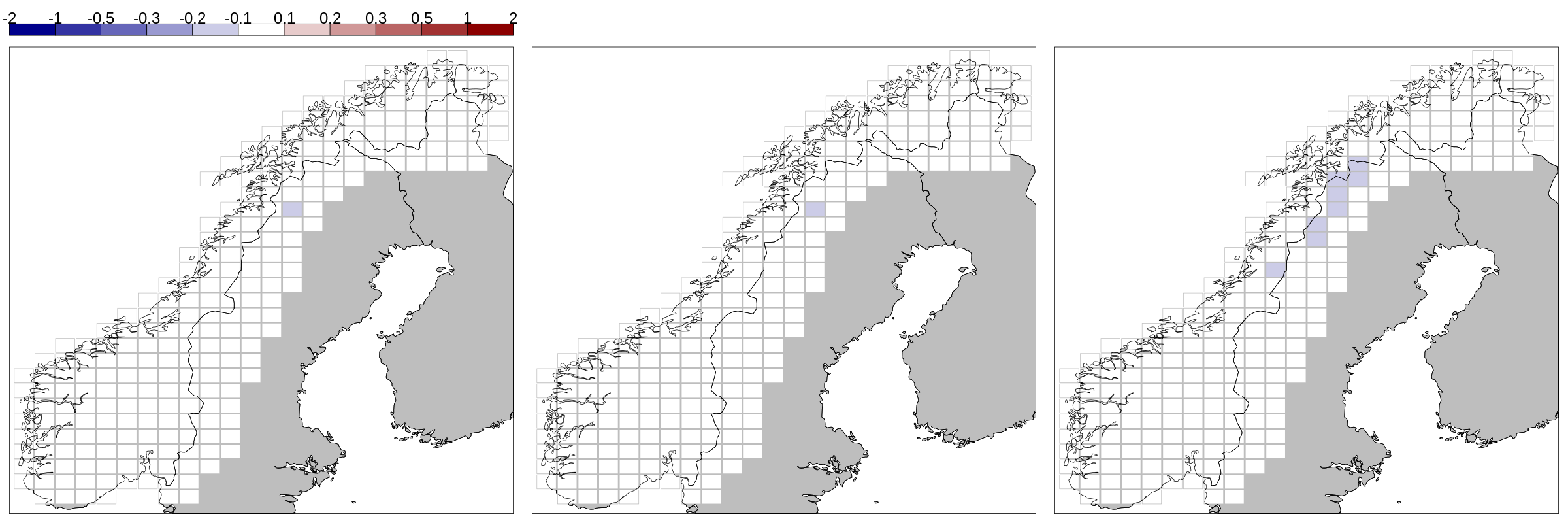}
    \caption{Same as Fig.~\ref{fig:res_tmx_djf} but for spring months: March, in the left panel; April, in the middle; May, in the right panel.}
    \label{fig:res_tmx_mam}
\end{figure}

\begin{figure}[h!]
    \includegraphics[width=\textwidth]{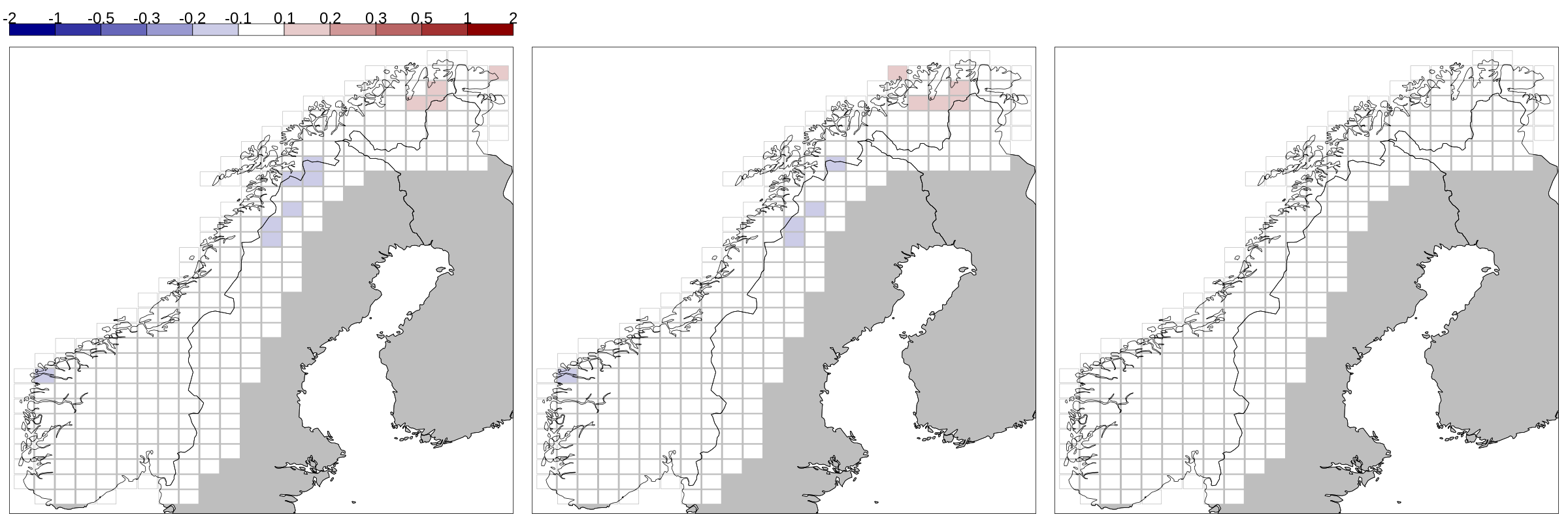}
    \caption{Same as Fig.~\ref{fig:res_tmx_djf} but for summer months: June, in the left panel; July, in the middle; August, in the right panel.}
    \label{fig:res_tmx_jja}
\end{figure}

\begin{figure}[h!]
    \includegraphics[width=\textwidth]{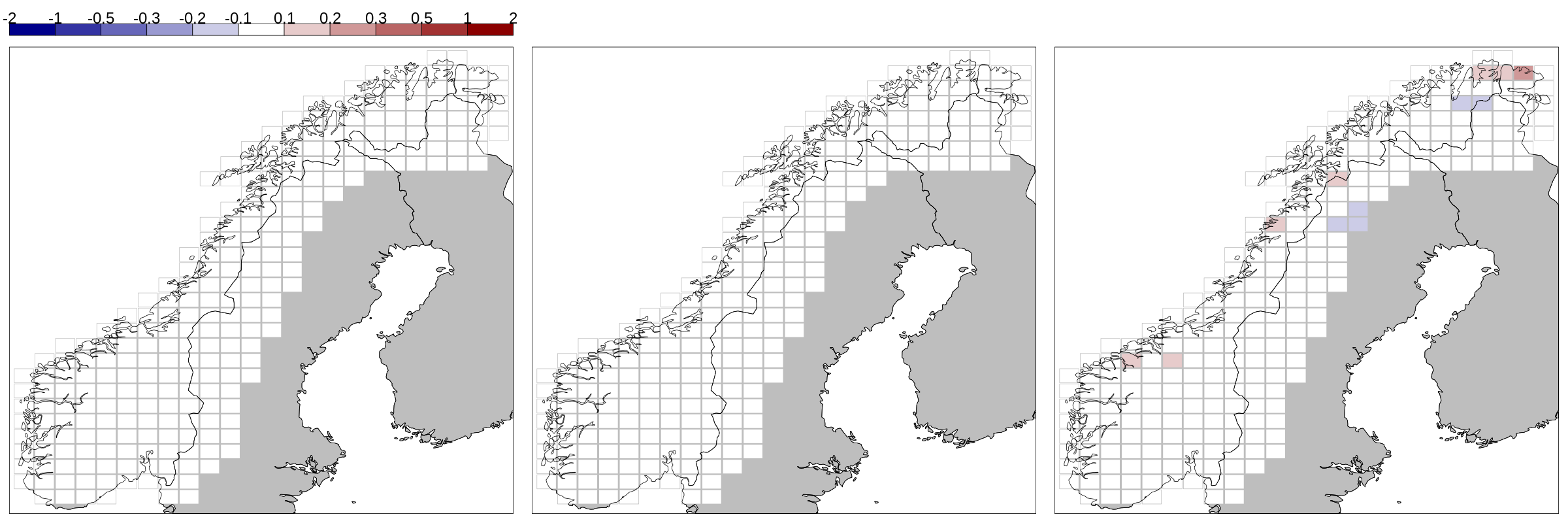}
    \caption{Same as Fig.~\ref{fig:res_tmx_djf} but for autumn months: September, in the left panel; October, in the middle; November, in the right panel.}
    \label{fig:res_tmx_son}
\end{figure}

\begin{figure}[h!]
    \includegraphics[width=\textwidth]{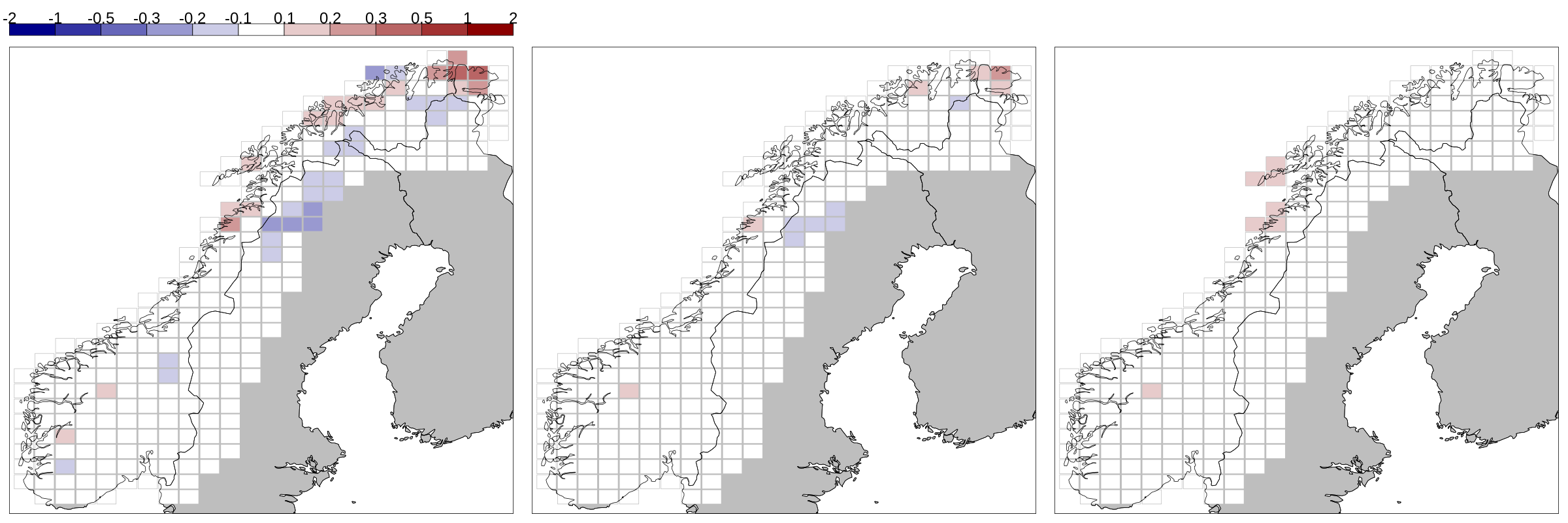}
    \caption{Same as Fig.~\ref{fig:res_tmn_djf} but for spring months: March, in the left panel; April, in the middle; May, in the right panel.}
    \label{fig:res_tmn_mam}
\end{figure}

\begin{figure}[h!]
    \includegraphics[width=\textwidth]{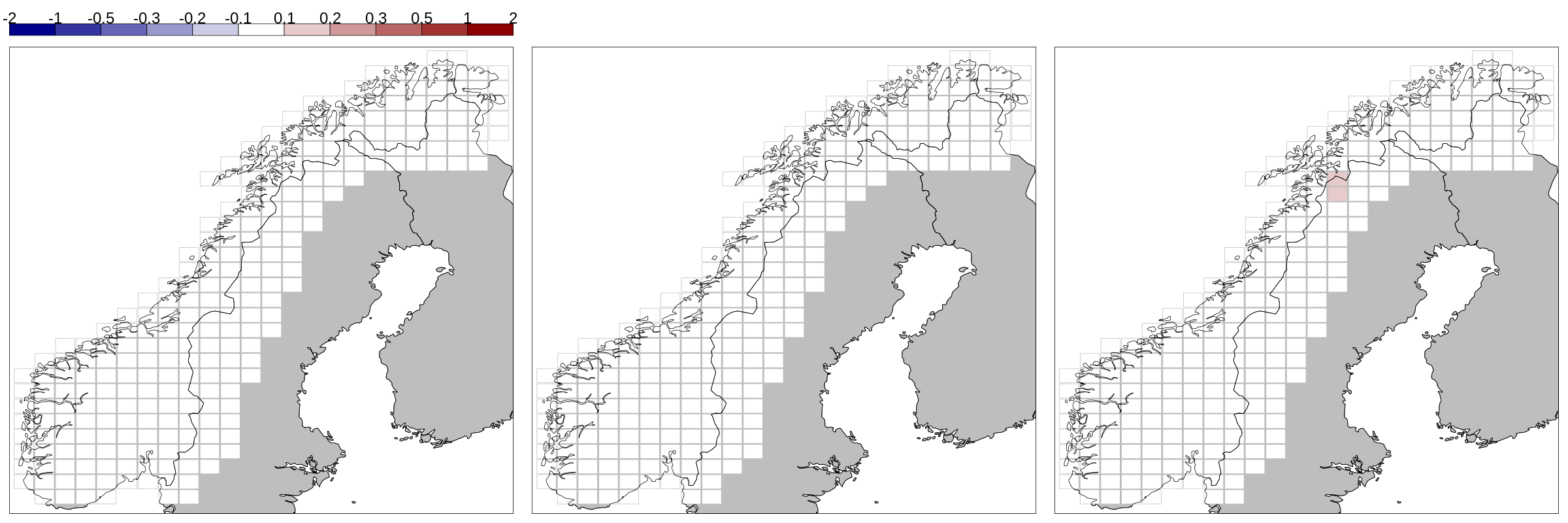}
    \caption{Same as Fig.~\ref{fig:res_tmn_djf} but for summer months: June, in the left panel; July, in the middle; August, in the right panel.}
    \label{fig:res_tmn_jja}
\end{figure}

\begin{figure}[h!]
    \includegraphics[width=\textwidth]{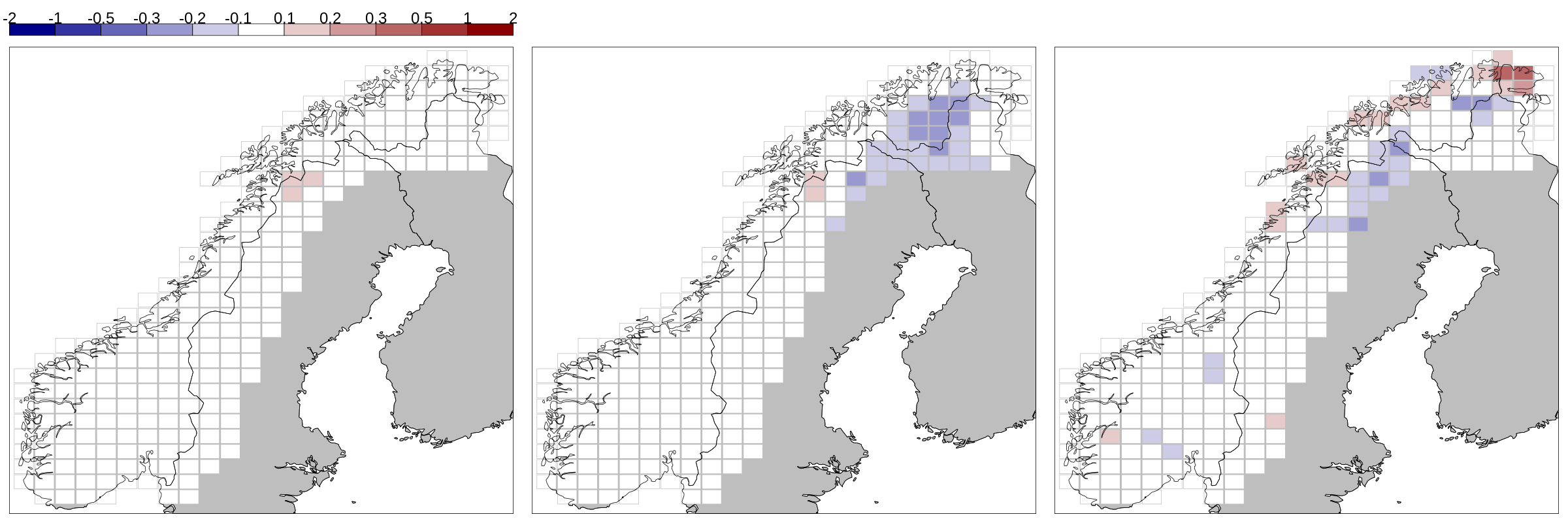}
    \caption{Same as Fig.~\ref{fig:res_tmn_djf} but for autumn months: September, in the left panel; October, in the middle; November, in the right panel.}
    \label{fig:res_tmn_son}
\end{figure}

The comparison of the daily minimum temperature is shown in Fig.~\ref{fig:res_tmn_djf} for winter months and Figs.~\ref{fig:res_tmn_mam}-~\ref{fig:res_tmn_son} for the other months.
The quantity considered is the monthly mean of daily maximum temperatures. The monthly mean is based on the period 1988-2017. Then, the differences between ver. 20.05 and ver. 18.12 are shown in the figures.
As for the other two temperature variables, the largest differences between the two versions are observed for the winter months, even though in the case of minimum temperature there are still significant differences in the northern part of the domain also for September, October and March.
The patterns of the deviations resemble also those observed for the daily mean temperature, with a region in northern Norway where ver. 20.05 is warmer (more than $1^\circ \mathrm{C}$) than ver. 18.12 and a region of the Scandinavian mountains in northern Sweden where ver. 20.05 is colder than ver. 18.12.
In addition, the deviations between minimum temperatures over South-Norway in winter are also significant for a few square boxes, up to $\pm 0.3^\circ\mathrm{C}$.

\section{Conclusions \label{sec:conclusions}}
\noindent seNorge\_2018 version 20.05 has been released and it is described in this document.
The observational gridded datasets include daily mean, minimum and maximum temperatures and daily total precipitation amounts over the Norwegian mainland for the time period 1957-2019.
The gridded fields are made available over a regular grid with 1 km of spacing between grid nodes in both easting and northing directions.

Version 20.05 is an update of ver. 18.12, which covers the same spatial domain but spans a smaller time period, from 1957 to 2017.
The data sources for the in-situ observations are the same for the two versions, however the observational datasets for ver. 20.05 have been collected from scratch, that is independently from ver. 18.12.
The in-situ observations have been quality controlled based on both visual inspection of the time series and the application of automatic procedures.
The set of in-situ observations used for ver. 20.05 is rather similar to the one used for ver. 18.12. A bit more observations have been used in ver. 20.05.

Version 20.05 includes some changes in the methods with respect to ver. 18.12.
In the case of precipitation, the reference fields used to scale the daily precipitation amounts in the statistical interpolation have been changed.
In the case of temperature, the statistical interpolation has been modified such that the pseudo-background, large-scale, field is adjusted locally on the basis of the surrounding observations over wider regions in ver. 20.05 than in ver. 18.12.

The combined impacts over the results of the updated input observational datasets and the modifications in the methods have been evaluated for the period 1988-2017 by considering monthly averaged deviations between ver. 20.05 and ver. 18.12.
Instead of considering grid points, we have aggregated deviations over square boxes with sides of 50 km.
In general, the deviations are rather small, especially in data dense regions, and this fact indicates a stability of the methods with respect to variations in some procedures.
However, in data sparse regions or where some new observations have been used in ver. 20.05, the differences between the two versions are evident even over 30-year averages.
For all variables, the largest differences occur during winter and they are more pronounced over few square boxes that are characterized by complex terrain and a sparse observational network. 
Future plans include a complete evaluation of seNorge\_2018 ver. 20.05 based on cross-validation to estimate the improvements in accuracy and precision with respect to ver. 18.12.


\clearpage
\section*{\hspace{17mm}Appendix: Data Access \label{sec:data_access}}
\noindent seNorge\_2018 version 20.05 datasets are available for public download under the Norwegian Licence for Open Government Data (NLOD, \url{https://data.norge.no/nlod/en/}). The data can be accessed from two distinct sources: Zenodo \url{zenodo.org} and MET's Norway THREDDS data server \url{thredds.met.no}.
A detailed description on how to access the data is available at the seNorge wiki-pages:
\begin{itemize}
    \item \url{https://github.com/metno/seNorge_docs/wiki/seNorge_2018}
\end{itemize}
In the wiki, it is also possible to get the latest news on the datasets and contact the developers by opening issues about specific topics.

The digital object identifiers (DOIs) of the datasets are:
\begin{itemize}
    \item TG, daily mean temperature  \citep{sN18_tg_20_05}. DOI is 10.5281/zenodo.3923706.
    \item RR, daily total precipitation amount  \citep{sN18_rr_20_05}. DOI is 10.5281/zenodo.3923703. In addition to RR, the archive also provides RRa (RR alternative), which is the field of daily total precipitation without the correction for the wind-induced undercatch.
    \item TX, daily maximum temperature  \citep{sN18_tx_20_05}. DOI is 10.5281/zenodo.3923700. In addition to TX, the archive also provides TXa (TX alternative), which is the field of daily maximum temperature without the check for consistency against TG and TN.
    \item TN, daily minimum temperature  \citep{sN18_tn_20_05}. DOI is 10.5281/zenodo.3923697. In addition to TN, the archive also provides TNa (TN alternative), which is the field of daily minimum temperature without the check for consistency against TG and TX.
\end{itemize}

\clearpage
\section*{\hspace{17mm}Supplementary Material}

\begin{table}[t]
\caption{Overview of variables and mathematical notation.
All the vectors are column vectors if not otherwise specified. 
If $\mathbf{X}$ is a matrix, $\mathbf{X}_i$ is its $i$th column (column vector) and $\mathbf{X}_{i,:}$ is its $i$th row (row vector).
If $\mathbf{x}$ is a column vector, $\mathbf{x}_j$ is its $j$th element.}
\label{tab:not}
\begin{tabular}{|l| l| l|}
\hline
symbol & description & dimension \\
\hline
$m$ & number of grid points & -  \\
$p$ & number of observations & -  \\
$l$ & index of OI configuration & $l=1,\ldots,4$  \\
$\boldsymbol{\lambda}$ & OI parameters &  not specified \\
$\mathbf{x}^{g,a} (\boldsymbol{\lambda})$ & TG analysis over the grid, obtained with OI parameters $\boldsymbol{\lambda}$  & m \\
$\mathbf{x}^{x,a} (\boldsymbol{\lambda})$ & TX analysis over the grid, obtained with OI parameters $\boldsymbol{\lambda}$  & m \\
$\mathbf{x}^{n,a} (\boldsymbol{\lambda})$ & TN analysis over the grid, obtained with OI parameters $\boldsymbol{\lambda}$  & m \\
$\mathbf{x}^{g,\mathrm{IDI}} (\boldsymbol{\lambda})$ & TG IDI over the grid, obtained with OI parameters $\boldsymbol{\lambda}$  & m \\
$\mathbf{x}^{x,\mathrm{IDI}} (\boldsymbol{\lambda})$ & TX IDI over the grid, obtained with OI parameters $\boldsymbol{\lambda}$  & m \\
$\mathbf{x}^{n,\mathrm{IDI}} (\boldsymbol{\lambda})$ & TN IDI over the grid, obtained with OI parameters $\boldsymbol{\lambda}$  & m \\
$\mathbf{w}^{l}$ & weight for OI with $\boldsymbol{\lambda}_l$, used in the merging & m \\
$\boldsymbol{\alpha}$ & normalization factor for the merging & m \\
\hline
\end{tabular}
\end{table}

\begin{algorithm}
\caption{seNorge\_2018 procedure for the production of daily gridded datasets for an arbitrary day.}
\label{a1}
\begin{algorithmic}

  \REQUIRE daily observations of temperature and precipitation collected from the data sources
  
  \REQUIRE daily averaged wind speed gridded field from numerical model output
  
  \REQUIRE Gridded fields of long-term monthly averages of precipitation from numerical model output
  
  \STATE{Production of 4 provisional versions of TG fields, OI with $\boldsymbol{\lambda}_l$, $l=1,\ldots,4$ }

  \STATE{Merging of the TG versions into the definitive TG field}

  \STATE{Adjust RR observations for wind-induced undercatch based on TG and the wind fields}
  
  \STATE{Production of RR gridded fields}
  
  \STATE{Production of 4 provisional versions of TX fields, OI with $\boldsymbol{\lambda}_l$, $l=1,\ldots,4$ }

  \STATE{Merging of the TX versions into the definitive TX field}
  
  \STATE{Production of 4 provisional versions of TN fields, OI with $\boldsymbol{\lambda}_l$, $l=1,\ldots,4$ }

  \STATE{Merging of the TN versions into the definitive TN field}

  \STATE{Check consistency between TX and TG. Modify TX where the consistency is violated}
  
  \STATE{Check consistency between TN and TG. Modify TN where the consistency is violated}
  
\end{algorithmic}
\end{algorithm}
 
\clearpage
\pagebreak

\bibliography{biblio}

\clearpage
\pagebreak

\end{document}